\documentclass[9pt,twocolumn,twoside]{IEEEtran}

\usepackage{soul,color}

\usepackage{times,enumerate}
\usepackage[usenames,dvipsnames,svgnames,table]{xcolor}

\usepackage{subfloat}

\usepackage{graphicx}
\usepackage{setspace}
\usepackage{bbm}
\usepackage{amssymb,latexsym,amsfonts,amsmath,cite,comment,relsize}
\usepackage[dvips]{psfrag}
\usepackage{amsthm}

\usepackage{adjustbox}

\usepackage{setspace}
\usepackage{url}
\usepackage{soul}
\usepackage{blindtext}
\usepackage{graphicx}
\usepackage{lipsum}

\usepackage{tikz}
\usepackage{pgfplots}
\pgfplotsset{every tick label/.append style={font=\small}}

\usepackage{relsize}
\usetikzlibrary{patterns}

\usepackage{epsfig}
\usepackage{subfig}
\usepackage{algorithm,algpseudocode}

\usepackage{setspace}

\newtheorem{theorem}{Theorem}
\newtheorem*{theorem*}{Theorem}
\newtheorem{lemma}{Lemma}

\newtheorem{proposition}{Proposition}
\newtheorem{corollary}{Corollary}

\theoremstyle{remark}

\theoremstyle{definition}
\newtheorem{definition}{Definition}

\newtheorem{assumption}{Assumption}
\theoremstyle{remark}
\newtheorem{example}{Example}
\newtheorem{remark}{Remark}

\newcommand{\BC}{\color{black}}

\tikzstyle{vertex}=[circle, draw, inner sep=0pt, minimum size=6pt]

\usepackage[sans]{dsfont}
\usepackage[font={small,it}]{caption}

\newcommand{\R}{\mathbb{R}}
\newcommand{\C}{\mathbb{C}}
\newcommand{\A}{\mathbf{A}}
\newcommand{\I}{\mathbf{I}}
\newcommand{\T}{\mathbf{T}}
\newcommand{\Z}{\mathbb{Z}}
\newcommand{\e}{\mathrm{e}}
\newcommand{\Sb}{\mathbf{S}}

\newcommand{\Tr}{\mathrm{Tr}}
\newcommand{\dm}{\mathfrak{d}(\A)}

\newcommand{\Sp}{\hspace{0.05cm}}

\newcommand{\suchthat}{\;\ifnum\currentgrouptype=16 \middle\fi|\;}

\title{
Space-Time Sampling for Network Observability  
}

\author{Hossein K. Mousavi$^{1}$, Qiyu Sun$^{2}$,  and Nader Motee$^{1}$
\thanks{$^{1}$The authors are with the Department of Mechanical Engineering and Mechanics, Lehigh University,
        Bethlehem, PA 18015, USA
        {\tt\small \{mousavi, motee\}@lehigh.edu} }%
        \thanks{$^{2}$ The author is with the Department of Mathematics, University of Central Florida, Orlando, FL  32816, USA
        {\tt\small qiyu.sun@ucf.edu}}%
}

\begin{document}

\maketitle
\thispagestyle{empty}
\pagestyle{empty}


\begin{abstract}
Designing sparse sampling strategies is one of the important components in having resilient estimation and control in networked systems as they make network design problems  more cost-effective due to their reduced sampling requirements and less fragile to where and when samples are collected. It is shown that under what conditions taking coarse samples from a network will contain the same amount of information as a more finer set of samples. Our goal is to estimate initial condition of linear time-invariant networks using a set of noisy measurements. The observability condition is reformulated as the frame condition, where one can easily trace location and time stamps of each sample. We compare  estimation quality of various sampling strategies using estimation measures, which depend on spectrum of the corresponding frame operators. Using properties of the minimal polynomial of the state matrix, deterministic and randomized methods are suggested to construct observability frames. Intrinsic tradeoffs assert that collecting samples from fewer subsystems dictates taking more samples (in average) per subsystem. Three  scalable algorithms are developed to generate sparse space-time sampling strategies with explicit error bounds.
\end{abstract}

\section{Introduction}
A common assumption in classical control systems for state estimation is that samples are collected periodically from some prescribed output sensors \cite{kalman1961new,anderson1979optimal,julier2004unscented}.  In practice, sampling strategies are designed subject to some given performance criteria and hardware/software constraints, e.g., achieving certain estimation quality, data processing power, battery-life of the sensors and processors, etc.  Although over-sampling may result in superior estimation quality, it is usually undesirable in networked systems that are equipped with spatially distributed sensors; examples include, spatially distributed networked robots, synchronous power networks, and platoon of self-driving vehicles. In these applications, designing sparse sampling strategies, that allow collecting samples aperiodically from only a fraction of subsystems, will reduce sensing costs due to the existing algorithmic, physical, hardware, and software constraints. These burdens are even more pronounced in networks with several thousands subsystems. Our goal in this paper is to propose a formal method to study properties and performance of various sampling strategies and devise scalable algorithms to design sparse sampling strategies in space and time with provable performance bounds.


There have been recent interest on revisiting notion of observability in the context of networked control systems. In \cite{bianchin2017observability}, the author revisit the notion of observability radius for a class of linear networks whose state matrices are adjacency matrix of some  weighted graphs. They provide conditions to verify whether such networks can preserve observability property in presence of structured (weighted) edge perturbations. The authors also suggest a heuristic algorithm to compute size of perturbations that result in loss of observability by finding their
smallest Frobenius norm. In \cite{pequito2015complexity} and \cite{pequito2016framework}, the problem of minimum constraint input selection are considered, where the objective is to find the smallest subset of inputs to ensure controllability. While it is shown that in general this problem is NP-hard, a subclass of such problems (by assuming dedicated inputs) can be solved efficiently with the aid of network graph algorithms. The author of \cite{olshevsky2014minimal} shows that the problem of approximating the minimum number of  input (output) variables to guarantee controllability (observability) is NP-hard. It is shown that one can find an efficient approximation of the problem by employing a greedy heuristic to select variables to maximize the rank increase of the controllability matrix. We refer to \cite{egerstedt2012interacting} for some related background and earlier works in this context. In \cite{tzoumas2017sensing}, the authors propose the problem of sensing-constrained LQG control, where contrary to the classical LQG, they look at the minimal sensing requirements for a desired control objective and tackle the problem by solving for suboptimal sensing strategies with by focusing only on finitely many sensing schedules.
They have also looked at  batch sensor-scheduling for linear-time invariant control systems \cite{tzoumas2016near}, where the goal is to design a sensing strategy to near-optimally estimate the concatenated states in a given finite horizon. {\BC In \cite{sinopoli2004kalman}, classic Kalman filtering problems were extended to the case with intermittent observations. } Our work is close in spirit to  \cite{jadbabaie2018deterministic, jadbabaie2018limitations}, where the authors investigate controllability of linear-time invariant networks and utilize randomized  algorithms for sparsification \cite{spielman2011graph} and (greedy) deterministic algorithms to obtain a sparse actuator scheduling. Furthermore, by allowing scaling in control inputs, they show that one can achieve desired levels of performance with respect to a class of performance measures. Prior to their work,  several authors had also considered problems related to sensor or actuator scheduling for control and estimation, for instance see  \cite{gupta2006stochastic, wu2013event,shi2011optimal,tanaka2015sdp} and the references therein. {\BC In another related work \cite{ranieri2014near}, the authors look at the sensor placement problem for optimal parameter estimation and  provide a near-optimal greedy algorithm for sensor selection.}

In this paper, our focus is on estimating the initial condition of a linear time-invariant (LTI) network from a set of state samples that are collected sparsely from a subset of subsystems aperiodically over some time interval. This problem is closely related to state observer design with sensing-constraints for linear dynamical networks. In Section \ref{sec:char}, we apply tools from (finite) frame theory to reformulate the network observability problem and show that one can extract an observability frame from any given set of samples that solves the observability problem. This key idea allows us to cast observability condition as whether a set of vectors forms a frame for the Euclidean space. This is particularly useful as every frame element is labeled by {\it where} and {\it when} it was taken. In Section \ref{sec:estimation_noise}, two types of measures, namely, standard deviation and differential entropy of the estimation error, are utilized to quantify quality of estimation for a given observability frame. They are also useful when one desires to compare estimation quality of various sampling strategies for a given network.
 We show that these estimation measures can be quantified using eigenvalues of the corresponding frame matrix. An important property of these estimation measures is that they are monotone with respect to the number of samples: by increasing the number of samples, the estimation measure does not deteriorate. In Section \ref{sec:construction}, we propose deterministic and randomized methods to generate observability frames for a given LTI network. It is shown that minimum required number of samples from each subsystem (location) depends on the degree of the minimal polynomial of the state matrix. We show in Section \ref{sec:funda-trade} that there are inherent fundamental limits on the best achievable levels of estimation quality, and intrinsic tradeoffs reveal an interplay between space-time samples: taking less samples (in average) per subsystem mandates collecting samples from more subsystems. In Section \ref{sec:sparse}, we discuss three methods for frame sparsification: (i) sparsification by leverage scores, which is developed based on notions of spectral graph sparsification  \cite{spielman2011graph}, (ii) random partitioning using Kadison-Singer paving solution \cite{marcus2013interlacing}, and (iii) greedy elimination using Sherman-Morrison rank-one update rule \cite{brookes2005matrix}. In all these algorithms, we obtain explicit error bounds for estimation-quality loss.  We assert that our bounds are rather conservative. The reason is that, contrary to the results of \cite{spielman2011graph,jadbabaie2018deterministic, jadbabaie2018limitations,siami2018network},  elements of a sparsified observability frame cannot be rescaled to compensate for estimation-quality loss. At the end, we support our theoretical finding by several simulation case studies.  This paper is an outgrowth of its conference version \cite{IFAC-MSM} and contains several new technical results, proofs, and simulation results. {\BC More specifically, we list some the main differences of the current work with \cite{IFAC-MSM} as follows. First, the omitted proofs from \cite{IFAC-MSM} have been added. Second, several results are new, including results in Sections \ref{sec:estimation_noise} and  \ref{sec:funda-trade}, Subsections \ref{subsec:timeshift} and \ref{subsec:greedy}, Theorem \ref{thm:tradeoffs}, and the running time analysis and corresponding numerical experiment. Third, several existing results have been improved, e.g., the minimal required number of samples per location in Theorems \ref{eq:perf_bound} and \ref{thm:Lyap}  have been refined and improved. Moreover, The counterpart of Theorem 14 in \cite{IFAC-MSM} (Theorem X) did not have a performance bound. In this work, we offer probabilistic guarantee for the outcome of sparsification. Forth, only a few sections from \cite{IFAC-MSM} have been transferred to this draft and that has been done after a thorough revision to enhance coherency of our presentation.  Some parts of \cite{IFAC-MSM} have been completely eliminated, e.g., space-time tradeoffs. Fifth, this paper covers a more compressive literature review. }

\section{Mathematical Notations }
The set of complex numbers, real numbers, nonnegative numbers, integers and nonnegative integers are shown by $\C,~\R,~\R_+,~\Z$ and $\Z_+$, respectively, and the imaginary number $\sqrt{-1}$ by $\mathrm{j}$.
For a given number $\gamma \in \mathbb{C}$, we define $\gamma \Z:=\big\{\gamma k~| ~k\in \Z \big\}$.
 For the $n$-dimensional Euclidean space $\R^n$, we denote its standard basis by $\{ e_1, \dots, e_n\}$
  and the inner product of  $x,y \in \R^n$ by $\langle x,y \rangle$.   For a vector $x\in \R^n$,  $\|x\|$ stands for  its Euclidean $2$-norm.
   For two families of vectors $\Phi_1$ and $\Phi_2$,  $\Phi_1 {\subset} \Phi_2$ implies that  $\phi \in \Phi_2$ for all  $\phi \in \Phi_1$.  We  use  the block capital  letters  to  denote  a  matrix  or  a  linear  operator, e.g., $\mathbf{X}$. The transpose of a matrix $\mathbf{X}$ is denoted by $\mathbf{X}^T$,  the matrix exponential of  a square  matrix $\mathbf{X}$  by  $\e^{\mathbf{X}}$, and the identity matrix of appropriate  size by $\I$.
 Eigenvalues of a positive semi-definite matrix $\mathbf{X} \in \R^{n\times n}$ is indexed in ascending order, i.e.,  $0 \leq \lambda_1(\mathbf{X})\leq \dots \leq \lambda_n(\mathbf{X})$; similarly, singular values of a square matrix  ${\mathbf X}$ are indexed from the smallest to the largest as $ 0 \leq \sigma_1(\mathbf{X})\leq\dots \leq \sigma_n(\mathbf{X})$; and the induced $2$-norm  is denoted by $\|\mathbf X \|=\sigma_n(\mathbf{X})$.
 Given two positive semi-definite
matrices ${\mathbf X}$ and ${\mathbf Y}$, we say that
${\mathbf X}\preceq{\mathbf Y}$ if ${\mathbf Y}-{\mathbf X}$ is positive semi-definite, and that
${\mathbf X}\prec{\mathbf Y}$ if ${\mathbf Y}-{\mathbf X}$ is positive definite.
A normal random variable with mean $\mu$ and covariance matrix $\mathbf{\Sigma}$ is denoted by $\mathcal{N}(\mu,\mathbf{\Sigma})$.  The expected value of a random variable is shown by $\mathbb{E}\{.\}$ and the probability of an event is denoted by $\mathbb{P}\{.\}$. The cumulative distribution function (cdf) of a scalar normal variable $\mathcal{N}(\mu,\sigma)$ is denoted by  $F(x;\mu,\sigma)$.  For  sequences $\{a_n\}_{n \geq 1}$ and $\{b_n\}_{n \geq 1}$ with positive elements,  notation $a_n=O(b_n)$ implies that $a_n/b_n$ is bounded.




 \section{Problem Statement}\label{sec:problem}

We consider linear dynamical networks that consist of multiple subsystems with state vector
\begin{align}
x:=[x_1,\dots,x_n]^T,
\end{align}
where $x_i \in \R$ is the state variable of subsystem $i \in \{1,2,\dots,n\}$. These subsystems are interconnected and their collective dynamics is governed by
\begin{align}\label{ds.def}
\dot x=\A x
\end{align}
in which $\A$ is time-invariant. It is assumed that initial state $x_0 \in \R^n$ of the network is unknown. In order to recover the initial state,  suppose that samples can only be collected from a subset of subsystems $\Omega=\{i_1,i_2,\dots,i_p\} \subset \{1,\dots,n\}$, where  $\Omega$ is called the set of sampling locations. At every spatial location $i \in \Omega$, sensors are allowed to take finite number of samples with different time stamps; the set of such sampling times is denoted by $\Theta_i$.  A sampling strategy for subsystem $i \in \Omega$ is given by the set of ordered pairs
\[
\mathfrak{S}_i = \big\{ (i,t) ~ \big| ~  t \in \Theta_i\big \}.
\]
A sampling strategy for the entire network can be obtained by
\begin{align*}
\mathfrak{S} = \bigcup_{i \in \Omega} \mathfrak{S}_i.
\end{align*}
For a given sampling strategy $\mathfrak{S}$, the corresponding  vector of samples or observations is shown by
\begin{align}\label{eq:observations}
y=\big[\Sp  x_i(t)+\xi_{i}(t) \Sp \big]_{ (i,t) \in \mathfrak{S}},
\end{align}
where measurement noises $\xi_i(t)$ in all samples are assumed to be independent from each other and have  normal (Gaussian) distributions with zero mean and $\sigma^2$ variance.  For a given set of sampling locations $\Omega$, the corresponding output matrix is defined by
\begin{equation}
\mathbf{C}_\Omega = \left[\begin{array}{ccc}e_{i_1}~| & \dots &|~e_{i_{p}}\end{array}\right]^T. \label{output-matrix}
\end{equation}

\begin{assumption}\label{assumption:obsv} The set of sampling locations $\Omega$ is chosen such that the pair $(\mathbf{A},\mathbf{C}_\Omega)$ is observable.
\end{assumption}

{\BC Verifying  observability of a  network  with respect to a given set of sampling locations is an interesting and active field of research on its own, and it is different from what we investigate here. For instance, in [24], the authors adapt a graphical approach to identify those sensors that are necessary for reconstruction of the initial state. Such results may offer option for $\mathbf{A}$ and $\Omega$ that satisfy Assumption 1.
}


The {\it research problem} of this paper is to characterize properties of sampling strategies that allow us to  recover initial state of linear network \eqref{ds.def} using sparse sets of samples in space and time.

\section{Characterization of Sampling Strategies} \label{sec:char}

We apply tools from finite frame theory to reformulate the observability problem and characterize its feasible sampling strategies.


\subsection{Reconstruction in Frame Theory}

The contents of this subsection are based on adjusted materials from reference \cite{casazza2013introduction}.

\begin{definition} For a given family of vectors $(\phi_i)_{i=1,\dots,m}$ in $\R^n$, the corresponding analysis operator $\T:~\R^n \rightarrow \R^m$ is defined by 
\begin{align}\label{eq:defT}
\T(x):=\left [\left \langle  x ,  \phi_i \right \rangle\right ]_{i=1,\dots,m}
\end{align}
and its frame operator $\Sb:~\R^n \rightarrow \R^n$  is defined by
\begin{align}
\Sb(x):= \sum_{i=1}^m \left \langle  x ,  \phi_i \right \rangle \phi_i.
\end{align}
\end{definition}
It is straightforward to verify that  operator $\T$ admits the following canonical  matrix  representation
\begin{align}
\T=[\phi_1 | \dots | \phi_m]^T \in \R^{m \times n}.
\end{align}
Thus, the canonical  matrix representation of the  frame  operator is
\begin{equation}\label{stt.eq}
{\bf S}={\bf T}^T {\bf T}  \ \in \R^{n \times n}.\end{equation}

\begin{definition} A family of vectors $(\phi_i)_{i=1,\dots,m}$ in $\R^n$ is a frame for $\R^n$ if there exists constants  $0 < \alpha\leq  \beta $ such that
$$
{\alpha} \|x\|_2^2 ~\leq~ \sum_{i=1}^m |\left \langle  x ,  \phi_i \right \rangle|^2 ~\leq~ {\beta} \|x\|_2^2 \text{ ~~for all }  x\in \R^n.
$$
The largest lower frame bound and smallest upper frame bound are called the optimal frame bounds.
\end{definition}


\begin{proposition} \label{frame.lem}
Let us consider a family of vectors $\Phi=(\phi_i)_{i=1,\dots,m}$ in $\R^n$. The following statement are equivalent:

\vspace{0.2cm}
\noindent (i) The family of vectors $\Phi$ forms a frame for $\R^n$.

\vspace{0.1cm}

\noindent (ii) The set of vectors  $\Phi$ span $\R^n$. Thus, $m=|\Phi| \geq n$.

\vspace{0.1cm}

\noindent (iii) The {corresponding} frame operator is positive definite, i.e.,  $\Sb \succ 0$, with optimal frame bounds
$
{\alpha}=\lambda_1({\bf S})$ and ${\beta}=\lambda_n({\bf S}).
$

\vspace{0.1cm}

\noindent (iv) The  corresponding analysis operator ${\bf T}$ is injective\footnote{i.e., its matrix representation has full column rank.} with a pseudo-inverse 
\begin{align}\label{eq:psuedo}
{\bf T}^{\dagger}:=\big ({\bf T}^T {\bf T}\big )^{-1}{\bf  T}^T= {\bf S}^{-1} {\bf T}^T,
\end{align}
which is  a left-inverse of $\T$ that satisfies ${\bf T}^{\dagger}{\bf T}={\bf I}$.
\end{proposition}

One of the well-studied  problems  in   frame  theory  is  to reconstruct an unknown vector $x\in \R^n$ from the following vector of observations
\begin{align} \label{eq:obsverations}
y=\T x=\left [\left \langle  x ,  \phi_i \right \rangle\right ]_{i=1,\dots,m} \in \R^m.
\end{align}
The following known result highlights role of $\T^{\dagger}$ in the reconstruction process  from these observations.

\begin{proposition} \label{eq:recovery} If the family of vectors $\Phi$ forms a frame for $\R^n$, then any vector $x\in \R^n$, with a corresponding vector of observations $y\in \R^m$ as in (\ref{eq:obsverations}), can be reconstructed via
\begin{equation}\label{reconstruction.eq0_0}
x={\bf T}^{\dagger} y,
\end{equation}
where $\T$ is the analysis operator of $\Phi$ and $\T^{\dagger}$ is given by \eqref{eq:psuedo}.
\end{proposition}


\subsection{Initial State Reconstruction}

The  solution  of the linear  network  (\ref{ds.def})  is  given by
\[x(t)=\e^{\A t}x_0,\]
where its  $i$'th component is
\begin{align} \label{eq:identify}
x_i(t)=e_i^T \e^{\A t}x_0=\big \langle  x_0 \ , \e^{\A^Tt} e_i  \big \rangle.
\end{align}
Based on the definition of a frame,  \eqref{eq:identify} reveals that
the following families of vectors  are  the only candidates for building constructors  to recover  initial state  of the network.
\begin{theorem} \label{eq:general_condition} Suppose that $\Omega$ is the set of sampling locations and $\Theta_i$ is the set of sampling times for each location $i\in \Omega$.
Every initial state of linear network  \eqref{ds.def}
 can be reconstructed from the set of  samples  that are collected according to sampling strategy $\mathfrak{S}=\{ (i,t)| ~i\in \Omega,~ t \in \Theta_i \}$ 
  if and only if  the family of vectors
\begin{align}\label{eq:phidef_general_thm}
\Phi(\A,\mathfrak{S}) \Sp=\Sp\left ( \e^{\A^Tt} e_i ~\big |\ ~ (i,t) \in \mathfrak{S} \right)
\end{align}
is a frame for $\R^n$.  
\end{theorem}

	
%


 The conclusion in Theorem \ref{eq:general_condition}
 asserts that initial state of the network can be recovered from the vector of observations
$$y=\T x_0=[x_i(t)]_{(i,t)  \in \mathfrak{S}}$$
using the following equation
$$
x_0=\T^{\dagger} y,
$$
where $\T$ is the analysis matrix of frame  \eqref{eq:phidef_general_thm}.

\begin{remark} A frame for $\R^n$ must contain at least $n$ vectors. Hence, the number of components in frame \eqref{eq:phidef_general_thm} satisfies
$$
\sum_{i\in \Omega} |\Theta_i| \geq n.
$$
This inequality implies that taking less spatial samples should be compensated by taking more temporal samples. This hints at an  inherent tradeoff between the minimum number of samples in space and time required for a successful initial state reconstruction.

\end{remark}



It turns out that observability at the sampling locations is a necessary condition for the reconstruction problem. 



{\begin{lemma} \label{lemma:necessary}
  Suppose that the family of vectors  \eqref{eq:phidef_general_thm} forms a frame for $\R^n$. Then, the pair $({\bf A},{\bf C}_\Omega)$ is observable.
\end{lemma}}

This can be interpreted as follows: if the sampling locations $\Omega$ create an unobservable output matrix $\mathbf{C}_\Omega$, then the initial state reconstruction will be always infeasible independent of the number of time samples.

In the rest of the paper, whenever it is not ambiguous, we drop argument of $\Phi(\A,\mathfrak{S})$ in \eqref{eq:phidef_general_thm} and simply write $\Phi$.
Whenever  \eqref{eq:phidef_general_thm} forms a frame, it will be referred to as an observability frame. The space of all observability frames in $\R^n$ is denoted by $\mathfrak{F}$.


\section{Estimation Measures }\label{sec:estimation_noise}

In the previous section, the reconstruction  problem was formulated in noise absence. One needs to solve an estimation problem when measurement noise is presented, which requires some appropriate mechanism to measure quality of the resulting estimations.  We start this section by showing that some useful estimation measures can be quantified in terms of  the frame eigenvalues (i.e., eigenvalues of the frame matrix) .

\subsection{ Estimation Measures}

Instead of pure measurements (\ref{eq:obsverations}), suppose  that  a noisy observation vector is collected
\begin{align} \label{eq:obsverations_noisy}
\hat y=y+\xi \in \R^m,
\end{align}
in which $\xi \in \R^m$ is a zero mean  Gaussian measurement noise with independent components and  covariance
$
\mathbb{E}  \left \{ \xi \xi^T  \right \}=\sigma^2 \I.
$
For  the linear network \eqref{ds.def}, the equation (\ref{eq:obsverations_noisy}) can be rewritten in the following form,
\begin{align}\label{eq:noisy_model}
\hat y=\T x_0+\xi,
\end{align}
where $\T$ is the analysis matrix associated with the observability frame $\Phi$  in \eqref{eq:phidef_general_thm}. Let us denote an estimation of $x_0$ by $\hat x_0$ and define the corresponding estimation error as
\begin{align}\label{eq:eta}
\eta:=\hat x_0-x_0.
\end{align}

\begin{definition}\label{def:monotone} An operator $\rho: \mathfrak{F} \rightarrow \R$ is called (decreasingly) monotone if $\rho(\Phi_2) \leq \rho(\Phi_1)$ for all $\Phi_1   \subseteq \Sp  \Phi_2$.
\end{definition}

In the following, we discuss two common estimation measures to compare different observability frames.

\vspace{0.2cm}
\noindent{{\it(i) Standard Deviation of the Estimation Error:}} For a given noisy observation vector (\ref{eq:noisy_model})  with underlying observability frame $\Phi$, this estimation measure is defined by
\[ \rho_d(\Phi) \Sp := \Sp \sqrt{\mathbb{E} \{ \| \eta  \|_2^2 \}}. \]
This  measure has been widely used to compute an optimal estimation via least-squares approximation \cite{goyal2001quantized}.


\begin{proposition} \label{prop:error} Suppose that a noisy observation vector $\hat y$ as in (\ref{eq:noisy_model}) is given. Then,
\begin{align}\label{eq:pseudo_estimation}
\hat x_0=\T^{\dagger} \hat y
\end{align}
is an unbiased estimator  for $x_0$ with $\mathbb{E}\{\hat x_0\}=x_0$ that minimizes $\| \T\hat x_0-\hat y\|_2$. Moreover, the (least-squares) estimation measure $\rho_d: \mathfrak{F} \rightarrow \R_+$ is monotone and can be characterized as
 \begin{align} \label{eq:perf_form_old}
\rho_d(\Phi)=\sigma \left (\sum_{i=1}^n \lambda_i(\Sb)^{-1} \right )^{1/2}
\end{align}
where 
$\lambda_1(\Sb),\dots, \lambda_n(\Sb)$ are  eigenvalues of the corresponding frame operator   ${\bf S}$.
\end{proposition}




%
%

\noindent{{\it(ii) Differential Entropy of the Estimation Error:} Since the measurement noise in (\ref{eq:noisy_model}) is assumed to be an independent Gaussian random variable $\mathcal{N}(0,\sigma^2 \I)$, one can use (\ref{eq:pseudo_estimation}) to show that the estimation error $\eta$ is also a normal random variable
 \begin{equation} \label{eq:g}
 \eta \sim \mathcal{N}(0,\sigma^2 {\bf S}^{-1}).
 \end{equation}
The differential entropy of  random variable $\eta \in \R^n$ with probability density  function $p(\eta)$ is defined as 
$$
{h}(\eta):=\int \limits_{\R^n} p(\eta) \log p(\eta)~d \eta.
$$
For Gaussian measurement noises, $h$ quantifies  the uncertainty volume of the estimation error.


 \begin{proposition}\label{prop:ent}
Under the Gaussian measurement noise assumption, the value of differential entropy of the estimation error is given by
\[ h(\eta) = \frac{1}{2} \rho_e(\Phi) + \frac{n}{2} \big(1+ \log(2 \pi  \sigma^2) \big) \]
with
\begin{equation}\label{entropy-measure}
\rho_e(\Phi) = -\sum_{i=1}^n \log \big(\lambda_i(\Sb) \big).
\end{equation}
Moreover, the above operator $\rho_e: \mathfrak{F} \rightarrow \R$ is monotone.
 \end{proposition}

\subsection{Effects of  Dwell-Time on Quality of Estimation }\label{subsec:timeshift}
Suppose that sensors are scheduled to take samples according to a sampling strategy $\mathfrak{S}$, but actual measurements are taken with a uniform dwell time $\delta \in \R$. Let us represent the resulting family of vectors by
\begin{align}\label{eq:phi_delta}
\Phi_\delta=\left(\left .  \e^{\A^T(t+\delta)}e_i  ~\right |~ (i,t)\in \mathfrak{S} \right ).
\end{align}
One can equivalently represent this set using \eqref{eq:phidef_general_thm} as
$$
\Phi_\delta=\Big( \Sp \mathbf{B}_\delta \Sp \phi~  \Big|~\phi \in \Phi \Sp \Big ),
$$
where $\mathbf{B}_\delta:=\e^{\A^T \delta}$ is full rank for all $\delta \in \R$. It is straightforward to verify that  elements of $\Phi_{\delta}$ span $\R^n$ if and only if  the elements of $\Phi$ span $\R^n$. Thus, the family of vectors $\Phi_\delta$ is a frame for $\R^n$ if and only if $\Phi$ forms a frame for $\R^n$. The next result shows that the estimation quality is not shift-invariant.


\begin{proposition}  \label{prop:performance_shifted}Suppose that measurement noise  (\ref{eq:obsverations_noisy}) has normal distribution   $\mathcal{N}(0,\sigma^2 \I)$. Then,   \begin{align}\label{eq:rho_shifted}
\rho_d(\Phi_\delta) ~\leq~ \sigma \Sp \Big( \Sp\sum_{i=1}^n \Sp { \sigma_i^2\big (\e^{-\A \delta}\big )  \Sp \lambda_i(\Sb)^{-1} } \Sp\Big)^{1/2}
\end{align}
and
\begin{align}
\rho_e(\Phi_{\delta})=\rho_e(\Phi)-\sum_{i=1}^n \log \Big( \Sp \Sp  \sigma_i^2\big (\e^{-\A \delta}\big ) \Big )  \Sp
\end{align}
in which $\sigma_i$'s are the singular values of the corresponding matrix.
\end{proposition}

The upper bound \eqref{eq:rho_shifted} becomes tight for $\delta=0$ because   $\sigma_i(\e^{\A\delta})=\sigma_i(\I)=1$ for all $i=1,\dots,n$.  When $\A$ is Hurwitz, according to inequality \eqref{eq:rho_shifted}, the estimation quality deteriorates as $\delta>0$ gets larger.  The reason is that magnitude of samples decrease and the measurement noise (with constant intensity) becomes more dominant as time goes by.  In fact,  \eqref{eq:rho_shifted} implies  that anti-stable state matrices neutralize negative effects of dwell time on the quality of estimation.

\section{Construction of Observability Frames} \label{sec:construction}






 Let us represent distinct eigenvalues of  state matrix $\A$ by  distinct eigenvalues $\lambda_1(\A), \dots, \lambda_q(\A)$ for some $q \leq n$ and its corresponding  {minimal} polynomial\footnote{The minimal polynomial of matrix $\A$  is the monic polynomial in $\A$ of smallest degree  such that $\mathrm{p}_{\A}(\A)=0.$ This should not be confused with the {characteristic} polynomial of a matrix, which is always of degree $n$ and only in certain cases coincides with the minimal polynomial \cite{burrow1973minimal}.} by
\begin{equation}\label{min-poly}
\mathrm{p}_{\A}(\lambda)\Sp = \Sp \prod_{m=1}^q \Sp \big(\lambda-\lambda_m(\A) \big)^{p_m}
\end{equation}
for some positive integers $p_m$, whose degree is denoted by $\dm$ which is less than or equal to $n$. To state our next result, we need to define the row vector map
\begin{equation}\label{Et.def}
       E(t):= \left [\e^{\lambda_m(\A) t} t^k\right ]_{\mathlarger{m=1,\dots, q }\atop \mathlarger{k=0,\dots, p_m-1}} \in \R^{1\times \dm}.
\end{equation}
\begin{theorem} \label{thm:output_random}
 Suppose that a sampling strategy $\mathfrak{S}=\{ (i,t)| ~i\in \Omega,~ t \in \Theta_i \}$ is adopted such that:

\noindent $\bullet$ At every $i \in \Omega$, $M_i:=|\Theta_i| \geq \dm$ samples are collected,  \\
$\bullet$ $\mathbf{E}_i$ has  full column rank, where
\begin{equation}\label{Econdition}
\mathbf{E}_i:=[E(t)]_{t\in \Theta_i} \in \R^{M_i \times \dm}.
\end{equation}
 Then, under Assumption \ref{assumption:obsv}, the family of vectors
 \begin{align}\label{eq:uniform_output}
{\Phi}=\left ( \left .\e^{\A^Tt} e_i~\right  |~(i,t) \in \mathfrak{S} \right )
\end{align}
forms a frame for $\R^n$.
\end{theorem}

Any frame for $\R^n$ must have at least $n$ vectors. Hence, prior to the application of Theorem \ref{thm:output_random}, a necessary condition for the total number of sampling times is
\begin{align}\label{eq:atleast}
|\mathfrak{S}|=\sum_{i \in \Omega} |\Theta_i| \geq n.
\end{align}
On the other hand, Theorem \ref{thm:output_random} requires $ |\Theta_i| \geq \dm$.  Comparing these two arguments implies that the resulting frame $\Phi$ from Theorem \ref{thm:output_random} will have  many redundant elements as the number of  locations $|\Omega|$ increases.  This motivates our investigation in the  next section to seek  scalable algorithms to construct sparse frames (in space and time) out of highly redundant observability frames.

According to Theorem \ref{thm:output_random}, the sufficient number  of samples at each location is $n$. This condition is rather conservative as it takes into accounts situations where samples are taken only from a very small (compared to $n$) subset of spatial locations. In Theorem \ref{theorem:statesampling}, it is shown that if $|\Omega|=n$, then we may collect as few as one sample from each spatial location.


The set of all time instances for which $\mathbf{E}_i$ is not full {column} rank have zero {Lebesgue} measure in the corresponding design space. In fact,  Theorem \ref{thm:output_random} suggests that one can comfortably skip rank verification step.


\begin{corollary} \label{cor:output_random}  For a given $\tau>0$, suppose that the sampling times in Theorem \ref{thm:output_random} are drawn randomly and independently from the uniform distribution over $[0,\tau]$. Then,  with probability $1$, the family of vectors in (\ref{eq:uniform_output}) is a frame for $\R^n$.
\end{corollary}

{\BC \begin{remark} Theorem \ref{thm:output_random} does not directly advise us to choose certain locations and times for an optimal estimation quality. { Nevertheless, the latter corollary motivates an approach for finding a sparse sampling strategy with an acceptable quality. First, we can randomly construct a rich and \emph{dense} set of space-time sampling indices. Then, we can use sparsification to discover the pivotal components of the sampling strategy (see next section)}. 
	\end{remark} }



\begin{example} \label{ex:simple}
Let us consider  the two-dimensional system
\begin{align}\label{eq:system_simple}
\dot x=\begin{bmatrix} 0 & -1 \\ 1 & 0\end{bmatrix} x.
\end{align}
We choose to sample only from the first subsystem; i.e., $\Omega=\{1\}$, $\mathbf{C}_\Omega=\begin{bmatrix} 1 & 0\end{bmatrix} $. Thus,  $(\A,\mathbf{C}_\Omega)$ is observable and  
\[
\e^{\A t}=\left [\begin{array}{ccc}
\cos(t) &~& -\sin(t) \\
\sin(t) &~&\cos(t)
\end{array}\right ].
\]
Let us pick sampling times $t_1,t_2 \in [0,\tau]$, i.e., $M_1=2$. The corresponding family of vectors is
\[
\Phi=\left (\e^{\A^Tt_1}e_1,\e^{\A^Tt_2}e_1\right )=\left ( \begin{bmatrix} \cos(t_1) \\ -\sin(t_1) \end{bmatrix},\begin{bmatrix} \cos(t_2) \\ -\sin(t_2) \end{bmatrix}  \right ).
\]
In this case, matrix (\ref{Econdition}) is
\[
\mathbf{E}_1= \begin{bmatrix} \e^{\mathrm{j}t_1} & \e^{-\mathrm{j}t_1} \\ \e^{\mathrm{j}t_2} & \e^{-\mathrm{j}t_2} \end{bmatrix} ~\Rightarrow ~\mathrm{det} (\mathbf{E}_1)=2\mathrm{j}\sin(t_1 - t_2).
\]
Hence, according to Theorem \ref{thm:output_random},  if $t_1-t_2 \neq k \pi $ for $k\in \Z$, then $\Phi$ is a frame.
 Alternatively, if we compute the frame matrix $\Sb$, using trigonometric identities, we get
\[
\mathrm{det} \left (  \Sb \right)=\sin^2(t_1 - t_2),
\]
which gives us the same constraints on the sampling times. Since the {Lebegues}  measure of the points  for which $\sin^2(t_1 - t_2)=0$ is indeed zero in $[0,\tau]\times [0,\tau]$,  any random  choices for $t_1$ and $t_2$ will result into a frame  with probability $1$. The latter observation agrees with Corollary \ref{cor:output_random}.  Next, we consider sampling $M_1=M$ samples at location $1$ for $M >2$. By induction on sampling times $\Theta_1=\{t_1,\dots,t_M\}$, we have
\begin{align} \label{eq:det}
\mathrm{det} \left (  \Sb \right)=\mathlarger{\sum}_{i=1,\dots,M \atop j=i+1,\dots,M}\sin^2(t_i - t_j).
\end{align}
Again, if $t_i-t_j \neq k \pi $ for $k\in \Z$, we get a frame out of these observations. Random sampling also results in a frame with probability $1$. Moreover,  $
\rho_d(\Phi)={\sqrt{2}\sigma}/{\sqrt{\mathrm{det}{(\Sb})}}$ and {it is shift-invariant}.
\end{example}

\vspace{0.2cm}

Now, we briefly look at periodic sampling strategy,\footnote{practical implication of this strategy is that sensors take
samples with some certain frequency based on a synchronized digital clock.} i.e., $\Theta_i=\{0,\delta,\dots,(M_i-1)\delta\}$ for every $i\in \Omega$, where   $\delta>0$ is a known sampling step-size.

\begin{theorem}\label{thm:periodic} Suppose that the sampling step-size satisfies
\begin{align}\label{lamb_cond}
\big(\lambda_m(\A)-\lambda_{m'}(\A) \big) \Sp \delta \not \in 2 \pi \mathrm{j} \Z
\end{align}
 for all  distinct eigenvalues $\lambda_m(\A)$ and $\lambda_{m'}(\A)$  of the state matrix $\A$. If $M_i\geq \dm$, then  the family of vectors
\begin{equation}\label{uniformframe_label}
{\Phi}= \Big( \Sp{\bf B}_\delta^{k}  e_i~\Big |~ i \in \Omega,~ k=0,\dots,M_i-1 \Sp \Big)
\end{equation}
forms a frame for $\R^n$, where ${\bf B}_\delta:=\e^{ {\bf A}^T\delta }$.
\end{theorem}

\begin{example}[Example \ref{ex:simple} continued] A sufficient condition for the sampling step-size for linear system \eqref{eq:system_simple}  is
$$
\mathrm{j}-(-\mathrm{j})\delta \Sp = \Sp 2 \mathrm{j} \delta  \Sp\not \in \Sp 2 \pi \mathrm{j}\ \Z ~\Rightarrow ~  \delta \not \in \pi  \Z.
$$
Alternatively, because $t_i-t_j=(i-j) \delta$, using the expression for $\mathrm{det}(\Sb)$ in  (\ref{eq:det}),  $\Phi$ is a frame if $\delta \not \in \pi  \Z$.
\end{example}

For a state matrix $\A$  whose all eigenvalues are real, one may verify that the requirement \eqref{uniformframe_label}
for any positive step-size $\delta$.  Therefore we have the following corollary by Theorem \ref{thm:periodic}.

\begin{corollary} If all eigenvalues of $\A$ are real, then for every step-size $\delta>0$ and sampling horizon $M_i\geq  \dm, i\in \Omega$, the family of vectors (\ref{uniformframe_label}) is a frame for $\R^n$.
\end{corollary}

Now, we consider the case where collecting  samples from all locations is possible and samples are taken in a small time range.

\begin{theorem} \label{theorem:statesampling} Suppose that $\Omega=\{1,\dots,n\}$ is the  set of sampling locations and  the set of sampling times $\Theta_i$, for each sampling location  $i\in \Omega$, is chosen such that $|\Theta_i|=M_i \geq 1$ and
\begin{align}\label{eq:time_condition}
\left [t^*,t^*+\delta^*\right ) \cap \Theta_i \neq \varnothing
\end{align}
for some $t^*\in \R,$ where   step-size $\delta^*>0$ is  given by
\begin{equation}\label{deltastar.def}
\delta^*<(\ln 2)\Sp \|\A\|^{-1}.
\end{equation}
Then, the sampling strategy \[\mathfrak{S}=\Big\{ (i,t)~\Big| ~i=1,\dots,n~\textrm{and}~ t \in \Theta_i \Big\}\] results in a  family of vectors
 \begin{align}\label{eq:nonuniform_complicated}
{\Phi}=& \left (\left .{\e^{\A^Tt} e_i}~\right |~(i,t) \in \mathfrak{S} \right )
\end{align}
that forms a frame for $\R^n$.
\end{theorem}


The time range $\delta^*$ in Theorem \ref{theorem:statesampling} only depends on the state matrix $\A$ and  is strictly positive.
Next, we consider the case where collecting  samples from all locations is possible and samples are taken randomly in a time range $[0, \tau]$, which is not necessarily in a small time range.

\begin{corollary} \label{cor:staterandom}Suppose that samples are collected from all subsystems, i.e., $\Omega=\{1,\dots,n\}$, at least once, i.e.,  $|\Theta_i| \geq 1$. Sampling times are drawn   randomly and independently  from the uniform distribution over  interval $[0,\tau]$.  Then, the resulting family of vectors $(\ref{eq:nonuniform_complicated})$ is a frame for $\R^n$ with probability $1$.
\end{corollary}

\begin{example}[Example \ref{ex:simple} continued] For the linear system (\ref{eq:system_simple}), let us consider a full state sampling with  strategy $\mathfrak{S}=\big\{ (1,t_1), (2,t_2) \big\}$ that results in vectors
\begin{align}\label{eq:ex_state}
\Phi=\left (\e^{\A^Tt_1}e_1,\e^{\A^Tt_2}e_2\right )=\left ( \begin{bmatrix} \cos(t_1) \\ -\sin(t_1) \end{bmatrix},\begin{bmatrix} \cos(t_2) \\ \sin(t_2) \end{bmatrix}  \right ).
\end{align}
For the corresponding frame matrix, we have
\begin{align}\label{eq:det_state}
\mathrm{det}(\Sb)=1 - \sin^2(t_1 - t_2).
\end{align}
Thus, $\Phi$ is a frame for $\R^2$ if and only if
\begin{align}\label{eq:cond_frame}
t_1-t_2 \neq \left (k+\dfrac{1}{2}\right ) \pi~~\textrm{for all} ~~k\in \Z.
\end{align}
Alternatively, $\delta^*$ in Theorem \ref{theorem:statesampling} satisfies
$$
\delta^*<\ln 2.
$$
According to  Theorem \ref{theorem:statesampling}, if sampling  times $t_1$ and $t_2$ satisfy
\begin{align}\label{eq:conserv_cond}
|t_1-t_2| <\ln 2, 
\end{align}
then the family of vectors \eqref{eq:ex_state} is a frame for $\R^n$. Comparing the two constraints characterized  by \eqref{eq:conserv_cond} and \eqref{eq:cond_frame} reveals that the resulting condition for sampling times from Theorem \ref{theorem:statesampling} is more conservative. To verify effectiveness of  Corollary \ref{cor:staterandom}, one can verify that the Lebesgue measure of all pairs of points  $\{t_1,t_2\}$ in $[0,\tau]^2\subset \R^2$ for which $\mathrm{det}(\Sb)=0$ is zero. As a result, one can randomly select sampling times $\{ t_1, t_2\}$   to construct a frame with probability $1$.
\end{example}
\vspace{0.1cm}

 \section{Frame Sparsification}\label{sec:sparse}
Suppose that a highly redundant set of samples from network \eqref{ds.def} is provided  for the estimation problem. This usually happens when a conservative sampling strategy $\mathfrak{S}$ is used and all subsystems are allowed to collect numerous samples over time. Even if the corresponding family of vectors
\begin{equation}
 \Phi = \left (\left .{\e^{\A^Tt} e_i}~\right |~(i,t) \in \mathfrak{S} \right )  \label{frame-redund}
\end{equation}
forms a frame, i.e., the resulting estimation problem is feasible, there are still unnecessary and undesired degrees of redundancy that should be trimmed away in order to enhance scalability properties of the estimation algorithms.

\begin{definition}
For a given design parameter $\theta >0$,  a family of vectors $\Phi_s$ in $\R^n$ is called a $\theta$-approximation of frame $\Phi$ if:

\vspace{0.1cm}
\noindent (i) $\Phi_s \subset \Phi$ and it has at most $\theta|\Omega|$  elements, \\
\noindent (ii) ${\Phi}_s$ is a frame for $\R^n$.
\end{definition}

For a given  (highly redundant) frame \eqref{frame-redund} and some parameter $\theta >0$,  our goal is to find a sampling strategy $\mathfrak{S}_s$ whose corresponding family of vectors
\[
\Phi_s=\left ( \left .\e^{\A^Tt} e_i ~\right |~ (i,t) \in \mathfrak{S}_s \right)
\]
is a $\theta-$approximation of \eqref{frame-redund}. Condition (i) mandates the sampling strategy $\mathfrak{S}_s$ to collect at most $\theta$ samples in average from all subsystems, i.e.,
\[
\bar{\Theta}(\mathfrak{S}_s) = \frac{1}{|\Omega|} \sum_{i \in \Omega} |\Theta_i(\mathfrak{S}_s)| = \frac{|\mathfrak{S}_s|}{|\Omega|}  \leq \frac{|\Omega| \theta}{|\Omega|} = \theta.
\]
Condition (ii) ensures the feasibility of the resulting estimation problem. In the following, we will discuss three methods to achieve our goal.

{\BC \begin{remark}
	Sparsity is a relative notion. In this paper, it is reasonable to consider a sampling strategy to be sparse if it takes almost linear number of samples (in terms of network size) in a given time window. This is the level of sparsity that the main result of this section, Theorem \ref{theorem:sparse}, achieves; we refer to next subsection. 
\end{remark}}



\subsection{Sparsification by Leverage Scores }

Our first approach is based on sparsification via the notion of effective resistances   \cite{spielman2011graph}, which
depends on the concentration properties of the sums of random outer-products and  was originally developed for sparsification of weighted graph Laplacians. Similar to  graph Laplacian, the frame matrix ${\bf S}$  is also a sum of  rank-one matrices
$$
{\bf S}=\sum_{\phi \in \Phi } \Sp \phi \phi^T.
$$
\begin{definition}
For a given finite frame $\Phi$, the leverage scores are  positive numbers that are defined by
 \begin{equation}\label{resistent.def}
 r_{\phi}(\Sb):=\phi^T {\bf S}^{-1} \phi
 \end{equation}
 for every $\phi \in \Phi$.
\end{definition}

One can associate a probability  mass  function $\pi: \Phi \rightarrow [0,1]$ to a given frame $\Phi$ using its leverage scores as follows:
\begin{equation}
\pi(\phi)  = \frac{r_{\phi}(\Sb)}{n}.\label{prob-elements}
\end{equation}
This gives a well-defined mass function as
\[
\sum_{\phi \in \Phi} \Sp \pi(\phi) \Sp =\Sp \frac{1}{n} \Sp \sum_{\phi \in \Phi}  \Sp  {\rm Tr} \Big ( {\bf S}^{-1} \phi \phi^T\Big )=1.
\]
As it is summarized in Algorithm \ref{alg:sparsification}, elements of $\Phi$ are sampled iteratively and independently, with replacement,  according to probability mass function \eqref{prob-elements}. A sampled element will be added to $\Phi_s$ if it is not already in $\Phi_s$. The resulting sparsified frame $\Phi_s$ will have at most $q$ elements. One may estimate $|\Omega_s|$, i.e., the number of sampling locations after sparsification, and obtain a reasonable  estimate for $\theta \leq q / |\Omega_s|$. Algorithm \ref{alg:sparsification} also assigns a weight to every  elements of $\Phi_s$ via weight function $w_s: \Phi \rightarrow \R_{+}$. Each execution of Algorithm \ref{alg:sparsification} returns a different realization of $w_s$, where $w_s(\phi)=0$ for $\phi \notin \Phi_s$.  These weights are useful in quantifying  estimation-quality loss due to  sparsification.  In fact, the weight function $w_s$ is a bounded random variable, where  $w_s(\phi)$
may assume different realizations drawn from  $\big\{ p (q \pi(\phi))^{-1}~\big|~p=0,1,\dots,q \big\}$.

\begin{algorithm}[t]
 \caption{Randomized Frame Sparsification}
 \label{alg:sparsification}
\begin{algorithmic}  
\algnewcommand\algorithmicinitz{\textbf{initialize:}}
\algnewcommand\Init{\item[\algorithmicinitz]}
 \renewcommand{\algorithmicrequire}{\textbf{input:}}
 \renewcommand{\algorithmicensure}{\textbf{output:}}
 \Require frame $\Phi=(\phi_1,\dots,\phi_{|\mathfrak{S}|})$ and design parameters $q,\epsilon>0$
\Ensure  set of vectors  $\Phi_s$ and weight function $w_s$ \vspace{0.15cm}
\Init  $\Phi_s=\varnothing$, $w_s(.) = 0$, $\Sb_s= {\bf 0}$
  \For {$k=1$ to $q$}
 \State sample an element from $\Phi$ with probability distribution $\pi$~$\rightarrow \phi$
 \State update weight function
 : $w_s(\phi)  \leftarrow w_s(\phi)  + (q\pi(\phi))^{-1}$
\vspace{0.05cm}
 \If {$\phi \notin \Phi_s$,}
\State add  $ \phi$ to $\Phi_s$
 \State update the frame matrix:
 $ \Sb_s \leftarrow \Sb_s+\phi \phi^T$
 \vspace{0.05cm}
 \EndIf
 \vspace{0.05cm}
  \EndFor
\end{algorithmic}
\end{algorithm}



 \begin{theorem} \label{theorem:sparse}
For a given frame $\Phi$ in $\R^n$, let us fix parameter $\epsilon \in (1/\sqrt{n},1]$ and the number of samples $q=O(n \log n/\epsilon^2)$. Then, the resulting set of elements $\Phi_s$ from  Algorithm  \ref{alg:sparsification} is also a frame for $\R^n$ with probability at least $1/2$. Furthermore, with probability at least $1/4$, the estimation-quality losses satisfy\footnote{Using monotonicity property of the estimation measures, one can show that upper bounds in \eqref{eq:comega_bound_1} and \eqref{eq:comega_bound_rhoe_1} are nonnegative. }
\begin{eqnarray}
\dfrac{\rho_d(\Phi_s)-\rho_d(\Phi)}{\rho_d(\Phi)}  &\leq &    -1+\sqrt{\dfrac{4\Sp\bar{\chi}}{1-\epsilon}} \label{eq:comega_bound_1} \\
{\rho_e(\Phi_s)-\rho_e(\Phi)} & \leq &   n \log \left ( \dfrac{4\Sp \bar{\chi}}{1-\epsilon} \right ), \label{eq:comega_bound_rhoe_1}
\end{eqnarray}
 where $\bar{\chi}:=\mathbb{E}\left\{ \chi\right\}$ and $\chi$ is  a random variable given by
 	\begin{align}\label{eq:omega_rv}
 	\chi:=\inf \Bigg\{{\gamma>0}~\Bigg|~     \sum_{\phi \in \Phi} w_s(\phi)  \big(\gamma - w_s(\phi) \big)\Sp \phi  \phi^T \succeq {\bf 0} \Bigg\}.
 	\end{align}	
 \end{theorem}


The backbone of this result is based on Theorem 1 of \cite{spielman2011graph} and asserts that  Algorithm \ref{alg:sparsification} trims off a given (highly redundant) frame and returns, with probability more than $0.5$, a new frame whose size is almost linear in network size. Moreover, it is shown that, with probability at least $0.25$, the estimation measures of the new (sparsified) frame stays within constant multiples/difference of the estimation measure of the original (redundant) frame.



\begin{corollary} \label{cor:wbound}
The random variable $\chi$, which is defined by \eqref{eq:omega_rv}, satisfies
\[
	\chi \Sp \leq \Sp \max_{\phi \in \Phi}\Sp w_s(\phi)
\]
almost surely.
Moreover,  under the settings of Theorem \ref{theorem:sparse}, inequalities
\begin{eqnarray}
\dfrac{\rho_d(\Phi_s)-\rho_d(\Phi)}{\rho_d(\Phi)}  &\leq &   -1+ \sqrt{\dfrac{4\Sp w_{\max}}{1-\epsilon}} \label{eq:comega_bound} \\
{\rho_e(\Phi_s)-\rho_e(\Phi)} & \leq &   n \log \left ( \dfrac{4\Sp w_{\max}}{1-\epsilon} \right ) \label{eq:comega_bound_rhoe}
\end{eqnarray}
holds with probability at least $1/4$, where  $$w_{\textrm{max}}:=\mathbb{E}\left\{  \Sp \displaystyle \max_{\phi \in \Phi}\Sp w_s(\phi) \Sp
\right\}.$$
\end{corollary}

{\BC	This corollary shows that there exists a clear relationship (even if it is not tight) between the performance loss and the magnitude of   parameter $w_{\max}$.}


The leverage scores disclose the importance of  every component with respect to the entire frame for the sake of estimation. For instance, if the network is asymptotically stable,  a component with a relatively large time label is expected to have a relatively small leverage score. Thus, such insignificant components are less likely to be sampled by Algorithm \ref{alg:sparsification} and can be trimmed off to achieve a comparable  estimation quality.

\vspace{0.1cm}

\noindent{{\it Running Time Analysis:}} Computing the inverse of $\Sb$ can be done in $O(n^3)$ operations, while computing the leverage scores using this matrix requires $O(|\Phi|n^2)$. We need to check for repeated samples, which does not increase the running time of the algorithm. Computing the frame matrix $\Sb_s$ can be done in $O(n\log n /\epsilon^2 \times n^2)=O(n^3 \log n/\epsilon^2)$ operations. Hence, the total running time of Algorithm \ref{alg:sparsification} is $O(n^3 \log n /\epsilon^2+|\Phi|n^2)$.

\subsection{ Random Partitioning and  Kadison-Singer Paving Solution}
If the leverage scores (\ref{resistent.def}) are uniformly bounded by  a small enough number, then components of a frame can be partitioned in a {balanced} manner in order to obtain two separate subframes with explicit bounds on their spectra. Such spectral bounds are useful to find bounds on the estimation quality of the resulting subframes. The next theorem is based on Corollary 1.3 of the recent seminal paper   \cite{marcus2013interlacing} that gives us a paving solution to the famous Kadison-Singer problem.

\begin{proposition} \label{prop:KS} Suppose that the leverage scores in \eqref{resistent.def} satisfy
\begin{equation}
 r_{\phi}(\Sb) \leq r^*  \label{bound-r}
\end{equation}
for all $\phi \in \Phi$ and some positive  number $r^* < 1.5 - \sqrt{2}$. Let us randomly partition $\Phi$ into two subfamilies $\Phi_1$ and $\Phi_2$ such that every element of $\Phi$, independent of others, belongs to either of the partitions with probability $1/2$.
Then, with a positive probability,  the resulting partition will satisfy
\begin{align*}
\begin{adjustbox}{max width=300pt}$
\left (1-\dfrac{(1+\sqrt{2r^*})^2}{2} \right )\Sb \preceq \sum \limits_{\phi \in \Phi_j} \phi \phi^T\preceq \left (1+\dfrac{(1+\sqrt{2r^*})^2}{2} \right )\Sb
$
\end{adjustbox}
\end{align*}
for $j=1$ and $2$.
\end{proposition}

For a given (highly redundant) observability frame,  the result of Proposition \ref{prop:KS} allows us to calculate the relative/absolute estimation-quality degradation of the resulting partitions.

\begin{theorem} \label{eq:perf_bound} Suppose that the leverage scores   \eqref{resistent.def} satisfy \eqref{bound-r}. Then, the randomly partitioned subfamilies  $\Phi_1$ and $\Phi_2$ from Proposition \ref{prop:KS} are both frames for $\R^n$ with a positive probability and the estimation-quality can be bounded as follows
\begin{eqnarray}
\dfrac{\rho_d(\Phi_j)-\rho_d(\Phi)}{\rho_d(\Phi)} & \leq & { \kappa}(r^*)  \label{eq:worstcase} \\
{\rho_e(\Phi_j)-\rho_e(\Phi)} & \leq & -n \log \left (1 - \frac{(\sqrt{2r^*} + 1)^2}{2}\right ) \label{eq:worstcase_e}
\end{eqnarray}
for $j=1,2$, where
\begin{align}
{\kappa(r)}:=\left (1 - \frac{(\sqrt{2r} + 1)^2}{2}\right )^{-1/2} - 1.
\end{align}
\end{theorem}

The quantity $\kappa(r^*)$ is a worst-case bound on the relative performance degradation of the  randomly partitioned subframes $\Phi_1$ and $\Phi_2$. This function has been illustrated in Fig. \ref{fig:mu_curve}. In simulations, we observe that comparably better bounds are achievable.
\begin{figure}[t]
\begin{center}
\includegraphics[trim = 11 0 10 10, clip,width=8cm]{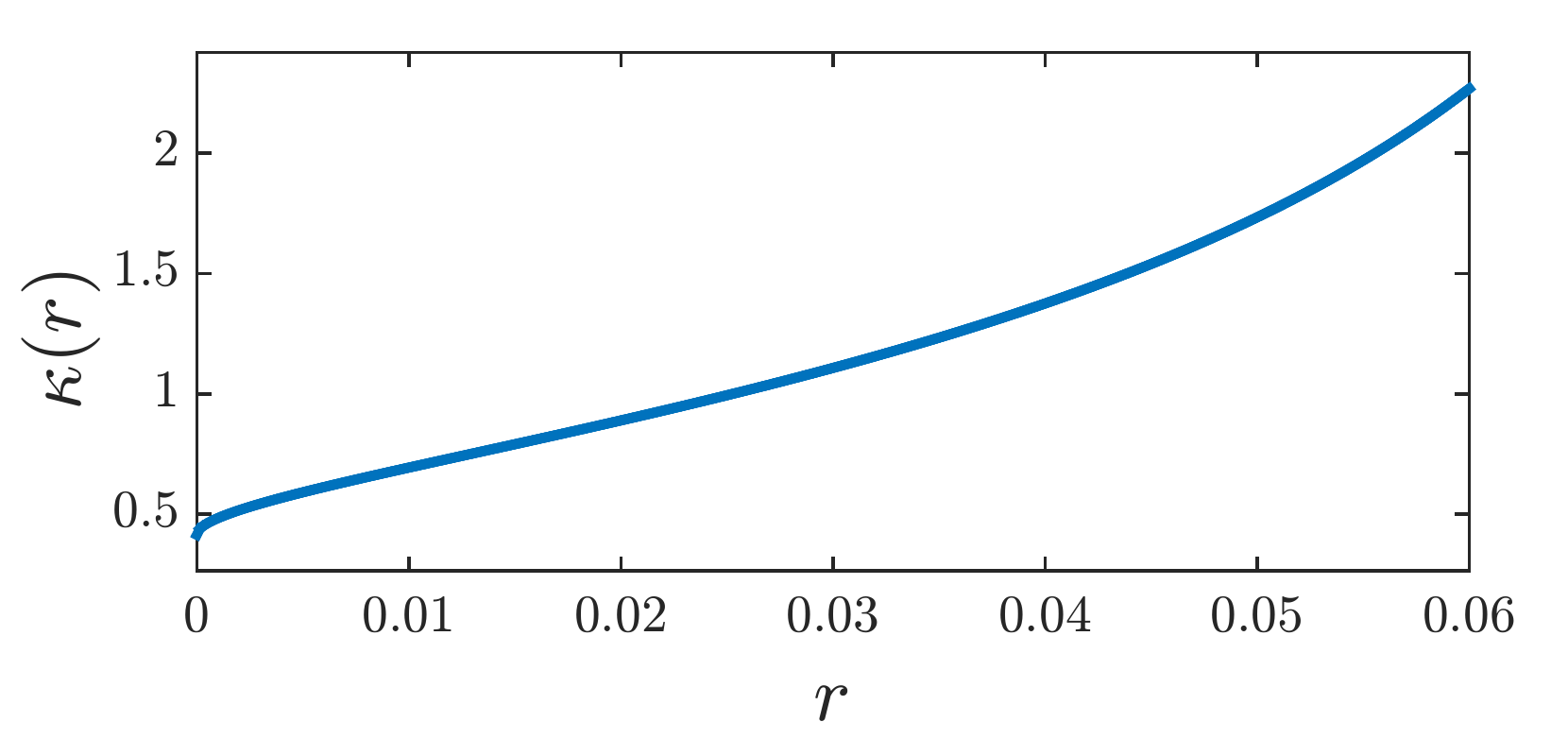}    
\caption{The worst case relative performance degradation based on the bound of Theorem \ref{eq:perf_bound}.  }
\label{fig:mu_curve}
\end{center}
\end{figure}

\vspace{0.1cm}
\noindent {\it Applicability of Random Partitioning:} When we deal with massive incoming samples (data) from the sensors, we expect the leverage scores to be small. In fact, we observe that the leverage scores satisfy
$$
\sum_{\phi \in \Phi} r_{\phi}(\Sb)=n ~\Rightarrow ~ \bar{r}(\Sb) = \dfrac{n}{|\Phi|},
$$
where $\bar{r}(.)$ stands for the average of the leverage scores. Hence, as a rule of thumb, one should expect that for fairly balanced observability frames with size
$$
|\Phi| > \big \lceil (6+4\sqrt{2})n \big\rceil,$$
the leverage score of a typical component in $\Phi$, in average, is less than $(6+4\sqrt{2})^{-1}=1.5-\sqrt{2}\approx 0.0858$.

\vspace{0.1cm}
\noindent{{\it Running Time Analysis:}}  Computing the frame matrix for each partition can be done in $O(|\Phi|  n^2)$. Hence, the random partitioning can be done in $O(|\Phi| n^2)$ operations.

\subsection{ Greedy Sparsification} \label{subsec:greedy}
In order to maintain a predetermined level of estimation quality and sparsity, one may consider using greedy algorithms that have been demonstrated to be useful in practice with satisfactory performance for a broad range of  combinatorial problems  \cite{karger1998random, 8070997}. In greedy frame sparsification, the core idea at every iteration is to eliminate that component of the frame which will  increase value of a given estimation measure less than the others.
At iteration $k$, let us denote the remaining frame and its frame matrix by  $\Phi_k$ and $\Sb_k$,  respectively. Eliminating a component $\phi$ from $\Phi_k$ corresponds to the following rank-one update
\begin{align}\label{eq:label}
\Sb_{k+1}=\Sb_k-\phi \phi^T
\end{align}
with $\Sb_0=\Sb$. According to the Sherman-Morrison formula   \cite{bartlett1951inverse}, one gets update rule
\begin{align}\label{eq:sinv_update}
\Sb_{k+1}^{-1}=\Sb_k^{-1}+\dfrac{\Sb_k^{-1} \phi\phi^T \Sb_k^{-1}}{1-\phi^T\Sb_k^{-1}\phi}.
\end{align}


\begin{algorithm}[t]
 \caption{Greedy Frame Sparsification}
 \label{alg:sparsification_greedy}
\begin{algorithmic}
\algnewcommand\algorithmicinitz{\textbf{initialize:}}
\algnewcommand\Init{\item[\algorithmicinitz]}
 \renewcommand{\algorithmicrequire}{\textbf{input:}}
 \renewcommand{\algorithmicensure}{\textbf{output:}}
 \Require frame $\Phi$ and its frame matrix $\Sb \succ 0$, $\mathfrak{s} \in (0,1)$, $\mathfrak{e}>0$
\Ensure  frame $\Phi_s$
\Init $\Sb_s=\Sb$, $\Phi_s=\Phi$  \vspace{0.2cm}
  \While {$\dfrac{\rho_{\Box}(\Phi_s)-\rho_{\Box}(\Phi)}{\rho_{\Box}(\Phi)} \Sp \leq \Sp \mathfrak{e}$ and $\dfrac{|\Phi_s|}{|\Phi|} \Sp \geq \Sp \mathfrak{s}$}
  \vspace{0.3cm}
 \State find minimizer $\phi_{\Box}^*$ via solving  \eqref{optim-1} or \eqref{optim-2}
  \State update the frame  matrix
 $$
\Sb_{s}^{-1}\leftarrow \Sb_s^{-1}+\dfrac{\Sb_s^{-1} \phi_{\Box}^*\phi_{\Box}^{*T} \Sb_s^{-1}}{1-\phi_{\Box}^{*T}\Sb_s^{-1}\phi_{\Box}^*}
$$
 \State update  $\Phi_s \leftarrow \Phi_s \backslash \phi_{\Box}^*$
 \vspace{0.2cm}
  \EndWhile
\end{algorithmic}
\end{algorithm}

\begin{proposition}\label{prop:updates} Upon eliminating a component $\phi$ from an observability frame $\Phi_k$, the estimation measures are updated according to
\begin{eqnarray}
\rho_d(\Phi_{k+1}) & = & \sqrt{\rho_d^2(\Phi_{k}) \Sp + \Sp \dfrac{\sigma^2  \Sp r_{\phi}(\Sb_k^2)}{1-r_{\phi}(\Sb_k)}} \label{eq:rho_update} \\
 \rho_e(\Phi_{k+1}) & = & \rho_e(\Phi_k)-\log\big(1-r_{\phi}(\Sb_k) \big). \label{eq:entropy_update}
\end{eqnarray}
\end{proposition}

At every iteration, the optimizer of $\rho_d(\Phi_{k+1})$ can be determined by solving the optimization problem
\begin{equation}
\phi_d^* = \arg \min_{\phi \in \Phi_k} ~ \frac{ \left\| \Sb_k^{-1} \phi \right \|^2}{1-\phi^T\Sb_k^{-1}\phi} \label{optim-1}
\end{equation}
and for $ \rho_e(\Phi_{k+1})$ by solving
\begin{equation}
\phi_e^* = \arg \min_{\phi \in \Phi_k} ~\phi^T\Sb_k^{-1}\phi.\label{optim-2}
\end{equation}

Algorithm \ref{alg:sparsification_greedy} details all necessary steps to compute a sparsfication of a (redundant) observability frame, where
we use notation $\Box \in \{d,e\}$. The algorithm stops whenever either a desired sparsity level $ \mathfrak{s} \in (0,1)$ or a maximum allowable relative estimation error $\mathfrak{e} >0$ has been achieved. This algorithm resembles the procedure of updating a performance measure of a linear consensus network when a new coupling link is added to the network \cite{siami2017growing}. {\BC The performance guarantees of the greedy methods in this context is a well-studied subject. In general, derivation of performance bounds heavily depends on the curvature conditions of the specific class of objective functions, e.g., sub-modularity, super-modularity or weak forms of these properties; please see  \cite{olshevsky2018non} and references in there.   }

\vspace{0.2cm}

\noindent{{\it Running Time Analysis:}} Computing  $\Sb^{-1}$ at the beginning requires $O(n^3)$. Then, at each iteration, one needs to compute and update the value of estimation measure for every vector in $\Phi$, which takes $O(|\Phi| n^2)$. Thus, in order to achieve  sparsity level $\mathfrak{s}$, one needs $O \big((1-\mathfrak{s})|\Phi|\times |\Phi| n^2 \big)=O \big((1-\mathfrak{s})|\Phi|^2n^2 \big)$ operations. Since $|\Phi| \geq n$, Algorithm \ref{alg:sparsification_greedy} can be implemented in $O \big((1-\mathfrak{s})|\Phi|^2 n^2 \big)$. Compared to running time of the randomized sparsification $O(n^3 \log n /\epsilon^2+|\Phi|n^2)$, the running time of the greedy method can be higher by an order of  $|\Phi|$.

\begin{remark}
Our proposed algorithms in this section employ some results from \cite{spielman2011graph,marcus2013interlacing}. The idea of sparsification via effective resistances are recently applied in controls community to obtain network abstraction as well as actuator scheduling in large-scale networked control systems \cite{jadbabaie2018deterministic} and \cite{siami2018network}.  The Kadison-Singer paving solution of \cite{marcus2013interlacing} is also utilized in \cite{jadbabaie2018deterministic} as a method of randomized actuator scheduling.
\end{remark}

\section{Fundamental Limits and Tradeoffs}\label{sec:funda-trade}


For a given linear network \eqref{ds.def} whose state vector is sampled based on an arbitrary  sampling strategy, we show that there are fundamental limits and tradeoffs on the best achievable values for the estimation measures and  space-time sparsity.


\begin{theorem} \label{thm:bounds}
	Suppose that state of the $n$-dimensional linear network \eqref{ds.def} is sampled under a sampling strategy $\mathfrak{S}$ with  total number of samples $|\mathfrak{S}|$. Then, the best achievable estimation measures are bounded from below by constants that are quantified by
	\begin{align}\label{eq:rho2_bound}
	\rho_d(\Phi) \geq  {\dfrac{\sigma n}{\nu \Sp \sqrt{|\mathfrak{S}|} }}
	\end{align}
	and
	\begin{align}\label{eq:bound_ent}
	\rho_e(\Phi) \geq n \log \left ( \dfrac{n}{\nu^2 \Sp |\mathfrak{S}|} \right),
	\end{align}
	where $\Phi$ is the observability frame  corresponding to $\mathfrak{S}$ and $\nu:=\nu(\A,\mathfrak{S})$ is defined by
	\begin{align}\label{eq:B_def}
	\nu(\A,\mathfrak{S})=\max_{ t \in \Theta_1 \cup \dots \cup \Theta_{|\Omega|} } \big\|\e^{\A t} \big\|. 
	\end{align}
\end{theorem}

The inequalities \eqref{eq:rho2_bound} and \eqref{eq:bound_ent}  give us some convenient rules of thumb about the scaling properties of the estimation measures. For instance,  if $|\mathfrak{S}| =O(n^2)$, then it can be deduced from \eqref{eq:rho2_bound} that\footnote{In the inequality \eqref{eq:rho2_bound}, the value of $\sigma/\nu$ can be interpreted as noise-to-signal ratio because  the value of $\nu$ relates to the norm of samples and $\sigma$ is the standard deviation of measurement noise. If $\A$ is Hurwitz, $\nu$ is always a finite number. On the other hand, for unstable networks, the noise-to-signal ratio loses its significance as sampling process is prolonged. As a consequence, the value of lower bounds in the inequalities \eqref{eq:rho2_bound} and \eqref{eq:bound_ent}  are usually small(er)  for unstable networks.}
$$\rho_d(\Phi) \geq  \frac{c\sigma}{\nu}$$
for some constant $c$. This  implies that the estimation quality cannot be enhanced beyond a hard limit. Such limitations are important in network design as they are independent of the network size and sampling strategy. For more discussions on significant role of fundamental limits in control, we refer to \cite{middleton1991trade,doyle2013feedback}.

\begin{theorem}\label{thm:Lyap}
	For a given linear network \eqref{ds.def}, let us assume that there exists $\delta >0$ such that all  distinct eigenvalues of its state matrix satisfy
	\begin{equation}
	\big(\lambda_i(\A)-\lambda_k(\A) \big) \Sp \delta \not \in 2 \pi \mathrm{j}\ \Z \label{condi-1}
	\end{equation}
	and it is sampled according to a sampling strategy with property\footnote{There is a practical implication for this assumption: sensors usually take samples with some certain frequency based on a synchronized digital clock.}  $\Theta_i \subset \delta \mathbb{Z}_+$ for all $i \in \Omega$. If $\A$ is Hurwitz, then universal (i.e., independent of number of samples) fundamental limits on the best achievable estimation measures emerge as follows
	\begin{eqnarray}
	\rho_d(\Phi) & \geq & \sigma \sqrt{\mathrm{Tr}(\mathbf{Q}^{-1})}, \label{ineq-1}\\
	\rho_e(\Phi) & \geq &  -\Tr \big(\log(\mathbf{Q}) \big) ,\label{ineq-2}
	\end{eqnarray}
	where  $\mathbf{Q} \succ 0$ is the observability Gramian, i.e., the unique solution of  Lyapunov equation
	\begin{align}
	\e^{\A^T \delta} \Sp \mathbf{Q} \Sp \e^{\A \delta} \Sp - \Sp \mathbf{Q} \Sp + \Sp \mathbf{C}_{\Omega}^T \Sp \mathbf{C}_{\Omega} \Sp = \Sp 0.
	\end{align}
	Also, the lower bounds can be achieved if and only if
	$\mathfrak{S} \Sp = \Sp \Omega \times \delta \mathbb{Z}_+.$
\end{theorem}

In an exponentially stable linear network, as time goes by, the magnitude of state dwindles compared to measurement noise. For such systems, Theorem \ref{thm:Lyap} predicts that estimation quality cannot be improved beyond some certain threshold even if the number of samples tends to infinity.


For a given sampling strategy $\mathfrak{S}$, the average number of samples per subsystem is quantified by
\begin{align}\label{eq:kappa_def}
\bar{\Theta}(
\mathfrak{S})
:=\dfrac{1}{|\Omega|}\sum_{i \in \Omega } |\Theta_i(\mathfrak{S})|.
\end{align}

\begin{theorem} \label{thm:tradeoffs}
	For given desired levels of estimation qualities $\rho_d^*, \rho_e^* >0 $ and parameter $\nu^* >0$, let us consider
	all $n$-dimensional linear networks \eqref{ds.def} whose pair of state matrices and sampling strategies belong to
	\[ \mathfrak{N}_{\Box}=\Big\{ (\A,\mathfrak{S})~\Big|~ \nu(\A,\mathfrak{S}) =\nu^*~~\textrm{and}~~\rho_{\Box}\big(\Phi(\A,\mathfrak{S}) \big) = \rho^*_{\Box}  \Big\}, \]
	where $ \Box \in \{d,e\}$. Intrinsic tradeoffs between the number of sampling locations and the average number of samples per subsystem transpire over $\mathfrak{N}_{\Box}$ that are characterized by
	\begin{align}\label{eq:rho3_bound}
	\bar{\Theta}(\mathfrak{S}) \cdot|\Omega|  ~ \geq~  \left({\dfrac{\sigma n}{\nu^* \Sp \rho_d^* }}\right)^2,
	\end{align}
	\begin{align}\label{eq:rho4_bound}
	\bar{\Theta}(\mathfrak{S}) \cdot|\Omega| ~\geq~ \dfrac{n}{\nu^{*2}} \exp \left ( -\dfrac{\rho_e^*}{n} \right).
	\end{align}
\end{theorem}

The result of Theorem \ref{thm:tradeoffs} asserts that intrinsic tradeoffs emerges among all linear networks with similar estimation quality and parameter \eqref{eq:B_def}: reducing number of sampling locations must be compensated by increasing the average number of samples per subsystem and vice versa.

\section{Numerical Simulations} \label{sec:num}

 Let us consider a linear dynamical network \eqref{ds.def} that consists of $n$ subsystems, which are  randomly and uniformly distributed over a square-shape spatial domain $[0,1] \times [0,1]$. The Euclidean (spatial) distance between the subsystems $i$ and $j$ is denoted by $\mathrm{dis}(i,j)$. If two subsystems lie in each others connectivity range, then there will be a coupling between the two subsystems and the corresponding entries in the state space will be nonzero numbers. More precisely, the state matrix $\A=[a_{ij}]$ is defined by
\begin{align}\label{aij}
a_{ij}=\left \{\begin{array}{lr} \zeta_{ij} \Sp \e^{-\mathfrak{a} \Sp \mathrm{dis}(i,j)^{\mathfrak{b}}}& \mathrm{dis}(i,j) \leq d
\\ 0 & \mathrm{dis}(i,j)>  d
 \end{array} \right..
\end{align}
for some $d >0$. The parameters $\mathfrak{a} >0$ and $0 < \mathfrak{b} <1$ determine decay rate of the couplings and spatial localization properties of the network. For instance, larger values of $\mathfrak{a}$ and $\mathfrak{b}$ result in more localized networks with short range couplings. To make our study  generic, the coefficients $\zeta_{ij}$ are independently and randomly chosen from $\mathcal{N}(0,1)$. In our simulations, we set $n=40$, $\mathfrak{a}=1$, $\mathfrak{b}=0.5$, and  $d=0.3$. We generate and save  one state matrix $\A$ that has both stable and unstable modes and use it in the following simulation studies.  The spatial locations of subsystems and their coupling topology are illustrated in Fig. \ref{fig:loc}.

\begin{figure}[t]
\begin{center}
\includegraphics[width=0.45\textwidth]{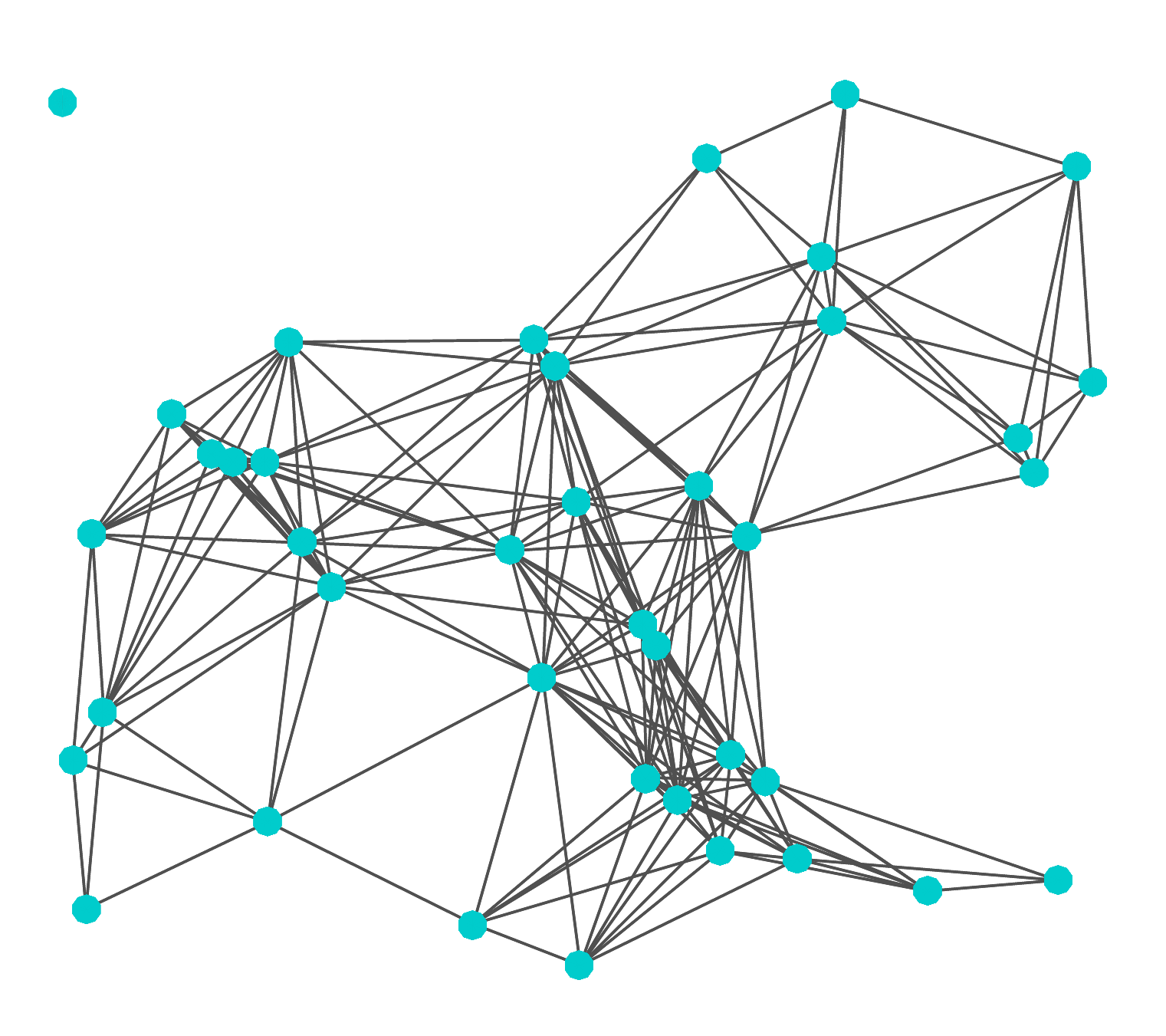}    
\caption{The spatial location of subsystems and their coupling structure.}
\label{fig:loc}
\end{center}
\end{figure}


\vspace{0.2cm}
\noindent{{\it Constructing Observability Frame:}} We set $\Omega=\{1,\dots,n\}$, $\tau=0.12$, and $M_i=M=44$ and utilize Corollary \ref{cor:output_random}  to construct an observability frame $\Phi$ with  format (\ref{eq:uniform_output}). The resulting frame contains $nM=1760$ vectors with a space-time representation illustrated by blue dots in Fig. \ref{fig:spacetime}.
\begin{figure*}[t]
\begin{centering}
\includegraphics[width=15.5cm]{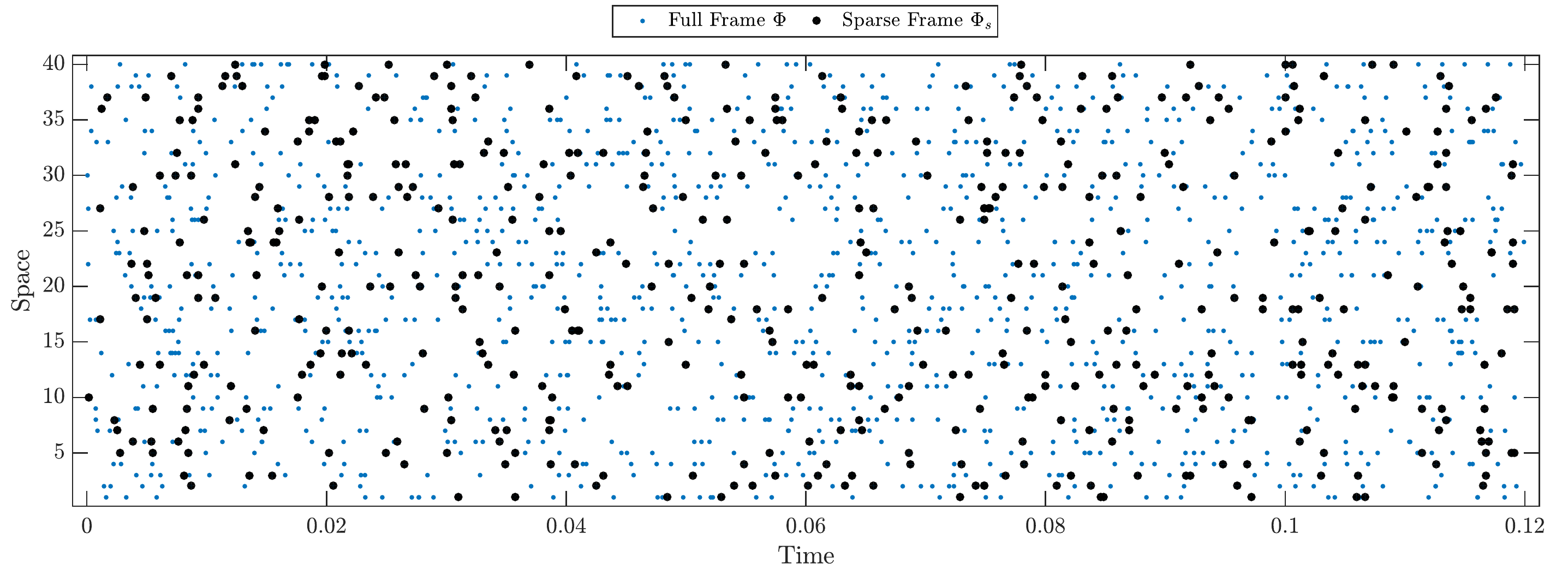}    
\caption{The space-time representation of the sampling strategies corresponding to frame $\Phi$ and  sparsified frame $\Phi_s$, with $|\Phi|=1760$ and $|\Phi_s|=503$.}
\label{fig:spacetime}
\end{centering}
\end{figure*}
For a noise intensity of $\sigma=0.1$, the value of the (least-squares) estimation measure is
$$
\rho_d(\Phi) \approx 0.0967.
$$

\noindent{{\it Randomized Frame Sparsification:}} For design parameters  $\epsilon=0.5$ and $q=590$, we apply Theorem \ref{theorem:sparse} and find  a sparsified frame with $|\Phi_s|=503$.
The spatial locations and time stamps of the sampling strategy corresponding to the sparsified  frame are illustrated by black circles in Fig. \ref{fig:spacetime}. The value of the estimation measure is
$$
\rho_d(\Phi_s)\approx 0.1883.
$$
The number of space-time samples has been reduced by $71\%$ for the price of $95\%$ relative estimation-quality loss.

\vspace{0.2cm}
\noindent{{\it Sparsification via Random Partitioning:}} Let us  reconsider the frame $\Phi$  shown in Fig. \ref{fig:spacetime}. By applying Proposition \ref{prop:KS}, random partitioning of $\Phi$ leads into two subsets $\Phi_1$ and $\Phi_2$. In our simulations, we repeat the random partitioning procedure  $5\times 10^5$ times. In each case, the minimum and maximum relative estimation-quality losses for  $\Phi_1$ and $\Phi_2$, i.e.,
$$\max_{j=1,2}\dfrac{\rho_d(\Phi_j)-\rho_d(\Phi)}{\rho_d(\Phi)}~~\textrm{and}~~\min_{j=1,2}\dfrac{\rho_d(\Phi_j)-\rho_d(\Phi)}{\rho_d(\Phi)}$$
 is computed and saved.  The histogram of this data is depicted in Fig. \ref{fig:h}. In this simulation, the minimum and maximum degradations are less than $0.55$ with a high probability. The theoretical estimate from Theorem \ref{eq:perf_bound} is
$$
r^*=\max_{\phi \in \Phi}~r_{\phi}(\Sb) \Sp\approx \Sp 0.0315\Rightarrow \kappa(r^*)\approx 1.1441.
$$
Our extensive simulations reveal that, in practice, one typically achieves comparably better estimation quality than our theoretical bounds (\ref{eq:worstcase}). In these simulations, the number of space-time samples are reduced by almost $50\%$ for the price of $55\%$ estimation-quality loss (in most outcomes of the simulations).

\begin{figure}
\begin{center}
\includegraphics[trim = 4 2 25 8, clip,width=8cm]{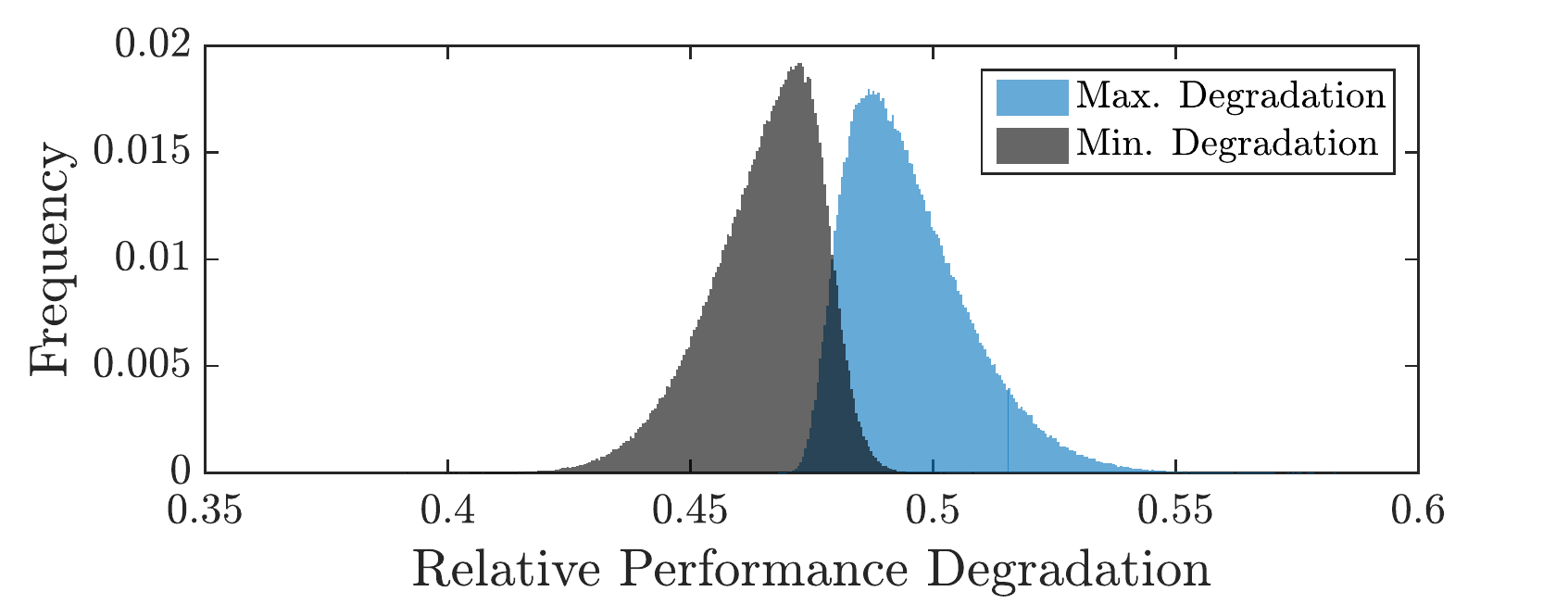}    
\caption{The histogram of the performance degradation after random partitioning.}
\label{fig:h}
\end{center}
\end{figure}

\begin{figure*}
\begin{center}
\includegraphics[width=16cm]{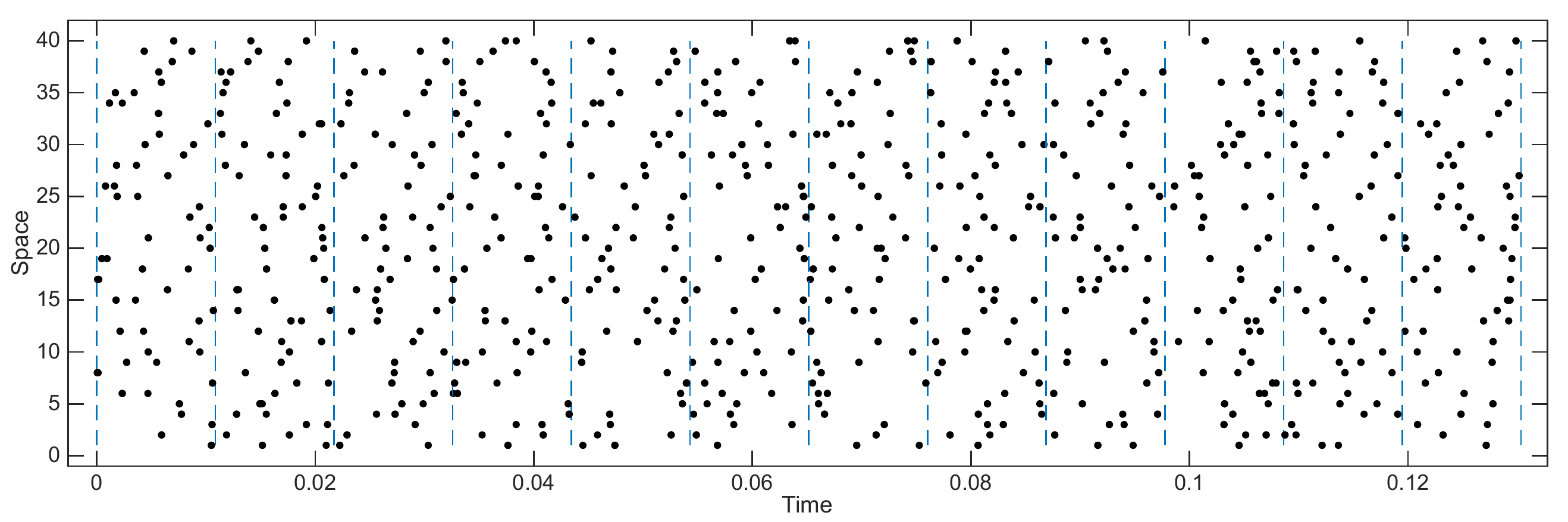}    
\caption{The space-time representation of the sampling points of the sequence of frames  $\Phi_1, \dots, \Phi_{12}$ that are separated by the dashed lines. }
\label{fig:spacetime_nu}
\end{center}
\end{figure*}

\vspace{0.2cm}
\noindent{{\it Performance of Randomized vs. Greedy Algorithms:}} In this simulation, we compare the estimation quality of the resulting sparsified frames from Algorithm \ref{alg:sparsification} and \ref{alg:sparsification_greedy}.  Using Algorithm \ref{alg:sparsification}, we construct  $25$ different sparsified frames by selecting  $25$ different values for  $\epsilon$ in $(1/\sqrt{n},1]$. We treat $q$ as a control parameter and vary its value between  $5902$ and $153$. For a fixed $q$, we compute the value of the estimation measure for all $25$ frames and save the one with the minimum value. When applying Algorithm \ref{alg:sparsification_greedy}, we change the desired sparsity level $\mathfrak{s}$ to get a sequence of sparsified frames. The outcome of our simulations is depicted in Fig. \ref{fig:compadre}, where one can observe that both methods result in almost similar estimation qualities. The only difference we can report is their running time  (on a personal computer with an Intel processor  using MATLAB): the randomized method (including $25$ experiments per $q$) took about  $1.68$ seconds, while the greedy method took $26.71$ seconds. This is consistent with our running time analysis for both algorithms.

\begin{figure}
\begin{center}
\includegraphics[width=8cm]{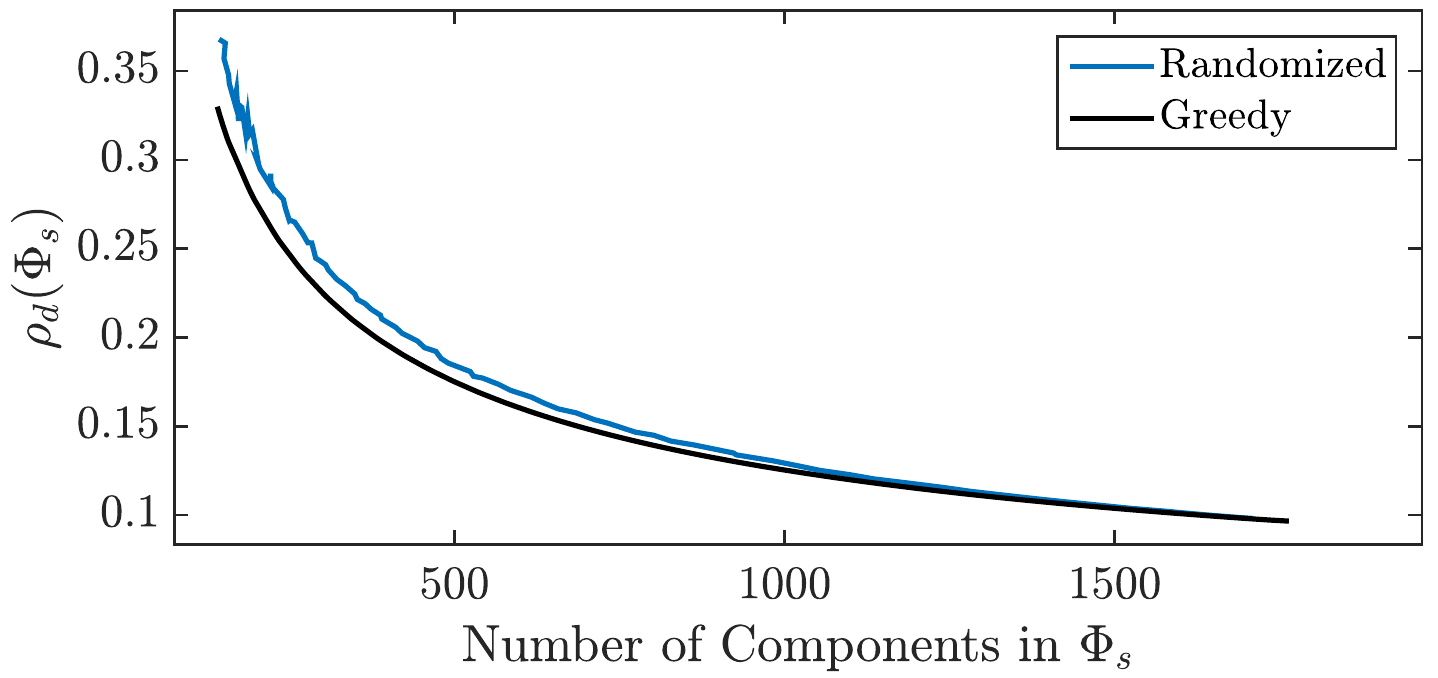}    
\caption{The estimation measure of the sparsified frames resulting from Algorithm \ref{alg:sparsification} (the randomized sparsification) and Algorithm \ref{alg:sparsification_greedy} (the greedy sparsification) are compared.}
\label{fig:compadre}
\end{center}
\end{figure}

\vspace{0.2cm}
\noindent{{\it Sequential Frame Construction:}}
We compute the value of $\delta^*$, which is defined in Theorem \ref{theorem:statesampling}, using the saved state matrix $\A$ and get $\delta^*\approx 0.0434$. We merge $N=12$ subframes in the time horizon, where $c_j=0.25$ is for every subframe $\Phi_j$ for $j=1,\dots,N$.  Sampling times $t_{ij}$ for each frame is chosen randomly and uniformly from time interval $[t_j,t_{j+1}]$. The space-time representation of these frames is illustrated in Fig. \ref{fig:spacetime_nu}. The resulting concatenated frame has $|\Phi|=Nn=480$ components with   estimation quality
$$
\rho_d(\Phi)\approx  0.1862.
$$

\noindent{{\it Estimation Quality Deterioration with Time Shifts:}} Let us consider the first observability frame $\Phi_1$ in Fig. \ref{fig:spacetime_nu} with $n$ components. First, we construct a family of observability frame $\Phi_\delta$, which is defined in \eqref{eq:phi_delta}, by increasing $\delta$ from $0$ to $0.5$. We compute exact value of the (least-squares) estimation measure for every $\Phi_\delta$. The result of our simulations is depicted in Fig. \ref{fig:shft} along with our theoretical upper bound \eqref{eq:rho_shifted}. One observes that our proposed upper bound is rather tight for all values of $\delta$ in $[0,0.5]$.


\begin{figure}
\begin{center}
\includegraphics[width=8.0cm]{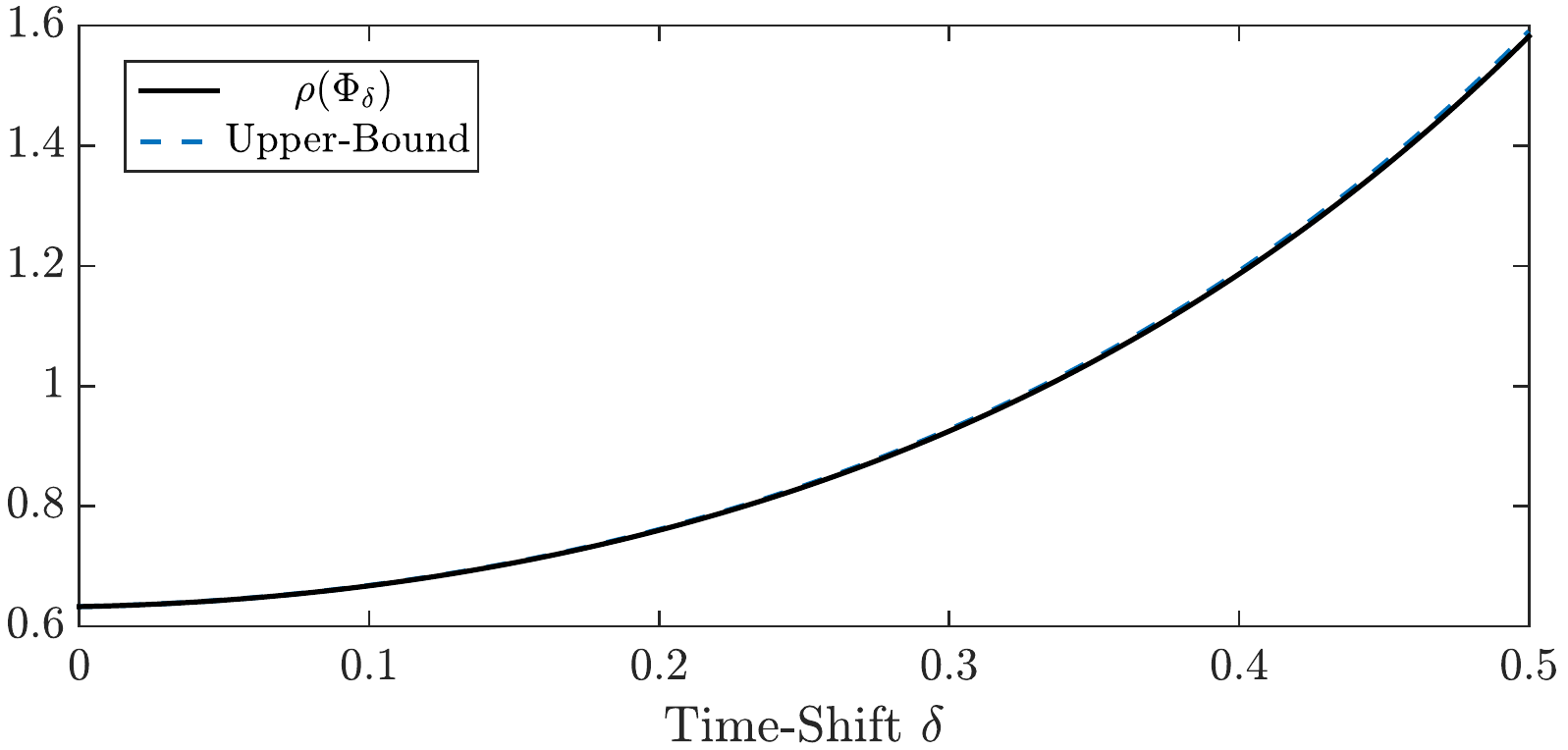}    
\caption{The estimation measure of the shifted frames is compared to our theoretical upper bound \eqref{eq:rho_shifted}.}
\label{fig:shft}
\end{center}
\end{figure}

 \section{Discussion and Conclusion}\label{sec:conc}

In this paper, we assume that  measurement noises in \eqref{eq:obsverations_noisy} are Gaussian and independent of each other.  When the measurement noises are dependent, the covariance of the estimation error \eqref{eq:eta}  will be
\begin{equation}\label{covar-eta} 
\mathbf{\Sigma}_{\eta}= 
\left (\T^T \mathbf{\Sigma}_{\xi}^{-1} \T \right )^{-1}
\end{equation}
where $\mathbf{\Sigma}_{\xi}$ is the covariance of the measurement noise. In general, we may not be able to expand \eqref{covar-eta} as a sum of rank-one matrices made of frame components. This was a useful property for  our developments in Section \ref{sec:sparse}. This case needs a thorough analysis which is beyond the scope of this paper. {\BC However, in the case that $\mathbf \Sigma_{\xi}$ is diagonal and with entries $$\mathbf \Sigma_{\xi}=\mathrm{diag}(\sigma_{i,t}^2)_{(i,t) \in \mathfrak{S}},$$   i.e., the observations are spatially and temporally independent, one can conduct a similar analysis by 
defining the frame components as 
$$
\Phi(\A,\mathfrak{S}) \Sp=\Sp\left ( \sigma_{i,t}^{-1}\, \e^{\A^Tt} e_i ~\big |\ ~ (i,t) \in \mathfrak{S} \right),
$$
where $\sigma_{i,t}^2$ is the variance of the observation error at $(i,t)$.  
}

Our results can be extended to include linear dynamical networks with arbitrary output matrices. In such networks, the sampling locations will be different from subsystem locations and the components of the observability frame \eqref{eq:phidef_general_thm} will take form $\e^{\A^T t} c^T$, where $c$ is a row of  output matrix. The main reason for working with output matrix \eqref{output-matrix} is to highlight inherent tradeoffs between minimum required number of samples in space and time in order to  achieve a certain estimation quality in linear dynamical networks; see the results of Section \ref{sec:funda-trade}.

In order to quantify estimation  quality, we consider two estimation measures: standard deviation and differential entropy of the estimation error. Depending on the specific design  criteria, one may choose another type of estimation  measure. For instance, if the reliability of the estimators is of significant importance for the network designer, one can utilize risk measures to evaluate the estimation quality. Two useful risk measures are:   risk of large aggregate deviations, i.e.,
\[
 \rho_{a} (\Phi):=\inf \Big\{\Delta \in \R_{+} ~\Big| ~\mathbb{P} \big\{ \|\eta\| \geq \Delta \big\} \Sp \leq \Sp \epsilon \Big\}
\]
or risk of large element-wise deviations, i.e.,
\begin{align*}
\rho_{r}(\Phi) :=\inf \Big\{\Delta \in \R_{+} ~\Big| ~\mathbb{P} \big\{ |\eta_i| \geq \Delta \big\} \Sp \leq \Sp \epsilon \Big\}.
\end{align*}
{\BC One can show }that these two risk measures are monotone according to Definition \ref{def:monotone} and as a result, one can effectively employ these measures instead of $\rho_d$ and $\rho_e$ in our proposed methodology.

The significance of fundamental limits and tradeoffs in Section  \ref{sec:funda-trade} is that they reveal what is achievable and what is not. This is practically plausible as it prevents us from searching for  sampling strategies with unachievable estimation qualities.

The results of Section \ref{sec:sparse} provide three methods to sparsify a given observability frame. Our theoretical error bounds are rather conservative. However, our extensive simulations assert that our proposed algorithms can achieve comparably better error bounds in practice.



\section*{Appendix I: Proof of The Remaining Results}
\label{appendix1}


\noindent{ \it Proof of Theorem \ref{eq:general_condition}:} 
		Let us assume that the family of vectors  \eqref{eq:phidef_general_thm} is a frame for $\R^n$ with analysis operator ${\bf T}$. From  Lemma \ref{eq:recovery}, initial condition can be recovered via
	\begin{equation}\label{reconstruction.eq0}
	x_0 = {\bf T}^{\dagger} y,
	\end{equation}
	where $y=\left [\left \langle  x ,  \phi_i \right \rangle\right ]_{\phi_i \in \Phi}$. In the next step, suppose that $(\ref{eq:phidef_general_thm})$ is not a frame for $\R^n$. Then,  the  frame matrix ${\bf S}$ will be singular by Proposition \ref{frame.lem}.  Thus, for every $x_1\in \R^n$,  two initial states $x_1$ and $x_1+x_2 $, in which   $x_2$ is a nonzero element of $\mathrm{null}({\bf S})$, produce the same vector of observation $y$. This is contrary to our assumption on unique determination of  the initial state of the network.

\noindent{\textit{Proof of Lemma \ref{lemma:necessary}:}}
Based on the Cayley-Hamilton theorem
\begin{align}\label{kale}
\e^{\A t}=\sum_{k=0}^{n-1} g_k (t) \A^k,
\end{align}
for some functions $g_k(t)$ for $k=0,\dots,n-1$. Using this fact, the analysis matrix corresponding to  the sampling strategy $\mathfrak{S}$ can be written as
\begin{align} \label{eq:T_CH}
\mathbf{T}=\left [\sum_{k=0}^{n-1} g_k (t)  \A^k e_i  \right ]_{(i,t) \in \mathfrak{S}}.
\end{align}
Assume that $\Phi$ is a frame but the pair $(\A,\mathbf{C}_\Omega)$ is not observable. Thus, the observability matrix
\begin{align}
O(\A,\mathbf{C}_\Omega)=\left [ \big(\A^k \mathbf{C}_\Omega^T\big)^T\right ]_{k=0,\dots,{n-1}}
\end{align}
has rank less than $n$. Because $\Phi$ is a frame,  rank of $\T$ is $n$. However, comparing  \eqref{eq:T_CH} with the form of the observability matrix, we observe that  rank of  matrix $\T$ in (\ref{eq:T_CH}) is at most equal to  rank of $O(\A,\mathbf{C}_\Omega)$. This is a contradiction, proving that $(\A,\mathbf{C}_\Omega)$ must be observable.

\vspace{0.1cm}
{\noindent{\it Proof of Proposition \ref{prop:error}:}
	The expression for the estimation measure  holds because we can write
	\begin{align*}
	\rho_d(\Phi)^2 &=\mathbb{E} \{ \| \eta  \|_2^2\}  = \mathbb{E} \Big \{\xi^T {\bf T}  {\bf S}^{-2} {\bf T}^T \xi\Big\}=
	\mathbb{E} \Big\{ {\rm Tr} \big(  {\bf S}^{-2} {\bf T}^T \xi \xi^T {\bf T}\big)\Big\}\\
	& =  \sigma^2 {\rm Tr} ({\bf S}^{-1})=\sigma^2\sum \limits_{i=1}^n \dfrac{1}{\lambda_i(\Sb)}.
	\end{align*}
	To see that $\rho_d$ is a monotone operator,  consider the following chain of observations:
	$$
	\Phi_1 \subseteq \Phi_2 \Rightarrow \Sb_1 \preceq \Sb_2 \Rightarrow  \Sb_2^{-1} \preceq \Sb_1^{-1} \Rightarrow \mathrm{Tr}(\Sb_2^{-1} )\leq \mathrm{Tr}( \Sb_1^{-1}).
	$$
	The last inequality implies that $\rho_d(\Phi_2) \leq \rho_d(\Phi_1)$.
}

\noindent{\it Proof of Proposition \ref{prop:ent}:}
For the random variable  $\eta$, using \eqref{eq:g}, we can show that (see Chapter 8 in \cite{cover2012elements})
\begin{align}
{h}(\eta)&=~ \dfrac{1}{2}\log \left ( \mathrm{det} (\sigma^2 \Sb^{-1}) \right )+\dfrac{n}{2}\log (2\pi e) \nonumber
\\  &= ~\dfrac{1}{2} \sum_{i=1}^n \log\left (\dfrac{\sigma^2}{\lambda_i(\Sb)}\right) +\dfrac{n}{2}\log (2\pi e)  \notag \\
&= \frac{1}{2} \rho_e(\Phi) + \frac{n}{2} \big(1+ \log(2 \pi  \sigma^2) \big).
\label{c1-c2}
\end{align}
To see that $\rho_e$ is a monotone operator,  consider the following chain of operations:
\begin{eqnarray*}
	\Phi_1   \subseteq  \Phi_2
	& \Rightarrow & \Sb_1 \preceq \Sb_2 \\
	& \Rightarrow  & \Sb_2^{-1} \preceq \Sb_1^{-1}  \\
	& \Rightarrow & \log(\Sb_2^{-1} )\preceq \log( \Sb_1^{-1}) \\
	& \Rightarrow &  \mathrm{Tr}( \log(\Sb_2^{-1} ) )\leq \mathrm{Tr}( \log(\Sb_1^{-1} )).
\end{eqnarray*}
The third inequality holds because $\log$(.), as a map from the cone of positive definite matrices to the set of symmetric matrices, is analytic and increasing on the cone of positive definite matrices.   The last inequality implies that $\rho_e(\Phi_2) \leq \rho_e(\Phi_1)$.

\noindent{\textit{Proof of Proposition \ref{prop:performance_shifted}:}}
Let us denote the analysis matrices corresponding to $\Phi$ and $\Phi_\delta$ by  $\T$ and $\T_\delta$, respectively.  It follows that
$$
\T_\delta=\T \e^{\A \delta}.
$$
Thus, the corresponding frame matrix for $\Phi_\delta$ is
$$
\Sb_\delta=\T_\delta^T\T_\delta=\e^{\A^T \delta}\T^T\T \e^{\A \delta}=\e^{\A^T \delta}\Sb \e^{\A \delta},
$$
where $\Sb$ is the frame matrix of $\Phi$.
As a result, we can see that
$$
\Sb_\delta^{-1}=  \e^{-\A \delta} \Sb^{-1} \e^{-\A^T\delta} .
$$
Therefore, estimation using the shifted frame $\Phi_\delta$  results in a normal error (random) variable
\begin{align}\label{eq:eta_shifted_2}
\eta \sim \mathcal{N}\left (0,\sigma^2 \Sb_{\delta}^{-1} \right )=\mathcal{N}\left (0,\sigma^2  \e^{-\A \delta} \Sb^{-1} \e^{-\A^T\delta} \right ).
\end{align}
In the next step, the estimation measures can be found as follows
\begin{eqnarray*}
	\frac{1}{\sigma^2} \Sp \rho_d(\Phi_\delta)^2 &=&  \mathrm{Tr} \left( \e^{-\A \delta}\Sb^{-1} \e^{-\A^T \delta} \right)\\
	& = &  \mathrm{Tr} \left( \e^{-\A^T \delta} \e^{-\A \delta}\Sb^{-1} \right) \\
	& = &  \mathrm{Tr}\left ( \big (\e^{\A \delta} \e^{\A^T \delta}\big )^{-1} \Sb^{-1}\right ) \\
	& \leq &
	\sum_{i=1}^n \sigma_i\left (   \big (\e^{\A \delta} \e^{\A^T \delta}\big )^{-1}   \right ) \Sp \sigma_i(\Sb^{-1}),
\end{eqnarray*}
where in the last equality we have used Von Neumann's trace inequality; we refer to  \cite{mirsky1975trace} for more details.
We can further write
\begin{eqnarray*}
	\rho_d(\Phi_\delta)^2 & \leq & \sigma^2 \sum_{i=1}^n \sigma_i^2 \big (   \e^{-\A \delta}   \big ) \Sp \lambda_i(\Sb^{-1})\\
	& =&  \sigma^2 \sum_{i=1}^n \dfrac{1}{\sigma_i^2(\e^{\A \delta}) \cdot \lambda_i(\Sb)}.
\end{eqnarray*}
For the last part of the proof, we have
\begin{eqnarray*}
	\rho_e(\Phi_\delta) & = & -\log (\det(\Sb_\delta))=-\log \big(\det(\e^{\A^T \delta}\Sb \e^{\A \delta})\big)\\
	& = & -\log \big(\det(\e^{\A^T \delta})\big) \det \big((\e^{\A \delta})\big)-\log \det (\Sb) \\
	& = & \rho_e(\Phi)-\sum_{i=1}^n \log \big(\lambda_i(\e^{\A^T \delta}\e^{\A \delta})\big) \\
	& = & \rho_e(\Phi)-\sum_{i=1}^n \log \big(\sigma_i^2(\e^{\A \delta})\big).
\end{eqnarray*}

%


\noindent{\it Proof of Theorem \ref{thm:output_random}: }
When ${\bf A}={\bf 0}$, the proof becomes  trivial as in this case $\e^{{\bf A}t} =\bf I$ for all $t\in \R$ and $\mathbf{C}_\Omega$ must be the identity matrix according to  Assumption \ref{assumption:obsv}.  When  ${\bf A}$ is nonzero, its minimal polynomial  has degree $\dm$ with  $1\le \dm\le n$.  Then, it follows that
\begin{equation}\label{appendixA.eq1}
\e^{\A t} \Sp =\Sp \sum_{k=0}^\infty \Sp \frac{t^k}{k!} \Sp {\bf A}^k \Sp =\Sp \sum_{k=0}^{\dm-1} \Sp g_k (t) \Sp {\bf A}^k
\end{equation}
for all $t\in \R$, where $g_k$'s are some  functions of time.
Let  ${\bf J}$ be the Jordan canonical form of  ${\bf A}$ and
\begin{equation}\label{appendixA.eq2}
{\bf A}= {\bf Q}^{-1} {\bf J} {\bf Q}.
\end{equation}
for some nonsingular matrix ${\bf Q}$.  By   combining \eqref{appendixA.eq1}  and \eqref{appendixA.eq2}, one obtains
\begin{equation}\label{appendixA.eq4}
\e^{{{\bf J}t}}=\sum_{k=0}^{\dm-1} \Sp g_k (t) \Sp {\bf J}^k
\end{equation}
for all $t\in \R$. Let us consider the minimal polynomial of $\A$ given by  \eqref{min-poly}, where $p_m$'s are some integer numbers for $1\le  m\le q$.
From the definition of a minimal polynomial, there are  Jordan blocks
${\bf J}_{\lambda_m, p_m}$
associated with every eigenvalue $\lambda_m(\A)$ whose  dimension  is $p_m$.
This together with \eqref{appendixA.eq4} implies that
\begin{equation*}  
\e^{{{\bf J}_{\lambda_m, p_m} t}} \Sp = \Sp \sum_{k=0}^{\dm-1} \Sp g_k (t) \Sp \left({\bf J}_{\lambda_m, p_m} \right)^k,
\end{equation*}
or equivalently,
\begin{equation}\label{appendixA.eq5}
\e^{\lambda_m(\A) t} \Sp \frac{ t^l}{l!}  \Sp = \Sp \sum_{k=l}^{\dm-1} \Sp g_k (t) \Sp {{k}\choose {l}} \Sp \lambda_m(\A)^{k-l}
\end{equation}
for all $0 \le l \le p_j-1$ and $t\in \R$.
The last equivalence holds as  $(j, j')$'th entry of $({\bf J}_{\lambda_m, p_m})^k$ with property $0\le j'-j\le \min(k, p_m-1)$ equals to ${{k}\choose {j'-j}} \lambda_m(\A)^{k-j+j'}$ and all other entries are equal to zero.
It is well known that functions $\e^{\lambda_m(\A) t} t^k$ for $m=1,\dots,q$ and $k=0,\dots, p_m-1$ for $t \in \R$ are linearly independent.
This, together with  \eqref{appendixA.eq5}, and the fact that $$\dm=\sum_{m=1}^q p_m$$ implies the existence
of a nonsingular matrix ${\bf D}$ that satisfies
\begin{equation}\label{appendixA.eq6}
G(t)= {\bf D} E(t),
\end{equation}
where $E(t)$ is defined in \eqref{Et.def}
and
$$G(t)=\big[g_k(t)\big]_{0\le k\le \dm-1}.$$
Let us denote $\Theta_i=\big\{t_{i,j}~\big|~ 1\le j\le M_i\big\}$. According to our assumptions, matrix ${\bf E_i}$ defined by \eqref{Econdition} has full row rank $\dm$.
This, together with \eqref{appendixA.eq1} and \eqref{appendixA.eq6}, implies the existence of scalars
$a_{i, j, k}$ such that
\begin{equation}\label{appendixA.eq7}
{\bf A}^k~=~  \sum_{j=1}^{M_i} ~a_{i, j, k}~ \e^{  {\bf A}^Tt_{i, j}}
\end{equation}
for $0\le k\le \dm-1$.  Let us denote
\[ \mathbf{C}_\Omega = \left[\begin{array}{ccc}e_{i_1}~\big| & \dots &\big|~e_{i_{p}}\end{array}\right]^T\]
and
$${\bf F}=\left[ ({\bf A}^T)^k \Sp e_i \right]_{i\in \Omega \atop 0\le k\le \dm-1}.$$
Based on Assumption \ref{assumption:obsv},
{\BC and the fact that $\dm$ is greater than or equal to the observability index of $(\mathbf A,\mathbf C)$, for $n \times \dm|\Omega|$ matrix ${\bf F}$ it holds that}
\begin{equation}
\label{appendixA.eq8}{\rm rank}({\bf F})=n,
\end{equation}
{\BC (for example see Section 6.3.1 in \cite{chen1998linear})}. 
Let us define
$$\T=\left [\e^{{\bf A}^Tt} \Sp e_i\right]^T_{ \mathlarger{i\in \Omega},~\mathlarger{t\in \Theta_i}}.$$
From \eqref{appendixA.eq7}, it follows that the
rank of matrix $\T$, whose size is $n\times |\mathfrak{S}|$, is
larger than or equal to rank of $\big[{\bf C}_\Omega {\bf A}^k \big]_{0\le k \le \dm-1}$. This together with \eqref{appendixA.eq8}
implies that $\T$ has rank $n$, which implies that the family of vectors \eqref{eq:uniform_output} form a frame for $\R^n$. 

\vspace{0.2cm}

\noindent{\textit{Proof of Corollary \ref{cor:output_random}:} }  First suppose that $|\Theta_i|=\dm$ for each $i \in \Omega$. Observe that  nonzero combinations  of  $\e^{\lambda_i(\A) t} t^k$ for $i=1,\dots,q$ and $k=0,\dots, p_i-1$  have only finitely many zeros.  This shows that for any $\tau>0$ and $i \in \Omega$
$$
\det({\mathbf{E}_i})=\mathrm{det}([E(t_j)]_{j=1,\dots,\dm}) \neq 0,
$$
for almost every choice of times $[t_1, \ldots, t_{\dm}]\in [0, \tau]^{\dm}$. Hence, if
we choose the sampling times $t_j \in \Theta_i$ with $|\Theta_i|=\dm$  randomly  in the range $[0, \tau]$ in an  independent  and uniform manner, then with probability one the requirement  \eqref{Econdition} is satisfied and $\mathbf{E}_i$ is full rank for each $i \in \Omega$.  Thus, by Theorem \ref{thm:output_random},  the resulting family of vectors is a frame with probability one.  If the number of random samples per location  $i \in \Omega$ increases beyond $\dm$, the result is still a frame. 

\vspace{0.2cm}

\noindent{\textit{Proof of Theorem \ref{thm:periodic}:}} First assume that for each location $i\in \Omega$, we choose $|\Theta_i|=\dm$. In this case, for every location $i \in \Omega$
$${\bf E}_i= [E(k\delta)]_{k=0,\dots, \dm-1}.$$
one observes that
$$
\det(\mathbf{E}_i)= C \prod_{1\le m<m'\le q} \left(\e^{\lambda_m(\A) \delta}-\e^{\lambda_{m'}(\A)\delta} \right)^{p_mp_{m'}}
$$
for some nonzero number $C$ depending only on $p_m$ with $1\le m\le q$. So the requirement \eqref{Econdition} is satisfied if
$$(\lambda_m(\A)-\lambda_{m'}(\A))\delta\not \in 2\pi \mathrm{j}\, \Z$$
for all distinct eigenvalues $\lambda_m(\A)$ and $ \lambda_{m'}(\A)$.  Therefore by  Theorem \ref{thm:output_random}, for $\mathbf{B}_\delta:=\e^{\A^T \delta}$, we know that
\begin{equation}\label{uniformframe}
{\Phi}':= \left (\left .{{\bf B}_\delta^{k}  e_i} \right |\ i \in \Omega,~ k=0,\dots,\dm-1 \right ).
\end{equation}
is  a frame. The family of vectors $\Phi$ given in the theorem satisfies ${\Phi}' \subseteq \Phi$. Thus, $\Phi$ is also a frame.

\noindent{\it Proof of Theorem \ref{theorem:statesampling}:} 
Since matrix $\e^{\A^Tt^*}$ is full rank for all $t^*\ge 0$,
one can verify that ${\Phi}$ in \eqref{eq:nonuniform_complicated} forms a
frame for $\R^n$ if and only if $\big({\e^{\A^T(t-t^*)} e_i}~\big |~(i,t) \in \mathfrak{S} \big )$
forms a
frame. Due to this shift-invariance property, we may safely assume that $t^*=0$ by shifting every element  in all $\Theta_i$'s by $-t^*$. Let us pick  $t_i\in \Theta_i$ for $1\le i\le n$. Then, for every  vector  $c=[c_1, \ldots, c_n]^T$, we have
\begin{eqnarray*}
	\left\|\sum_{i=1}^n  c_i e^{\A t_i} e_i- \sum_{i=1}^n c_i e_i\right\|
	&= & \left\| \sum_{m=1}^\infty \frac{\A^m}{m!}  \sum_{i=1}^n c_i  t_i^m e_i\right\|\\
	&\le & \sum_{m=1}^\infty \frac{\|\A\|^m}{m!} ~\left\|\sum_{i=1}^n c_i  t_i^m e_i\right\|\\
	&\le  & \sum_{m=1}^\infty \frac{\|\A\|^m (\delta^*)^m}{m!} \|c\| \\
	& = & \left(e^{\|\A\| \delta^*}-1 \right)\Sp \|c\|.
\end{eqnarray*}
This implies that
\begin{equation*}
\left(2-e^{\|\A\| \delta^*} \right) \Sp \|c\| \Sp \le  \Sp \Big\|\sum_{i=1}^n  c_i e^{\A t_i} e_i\Big\| \Sp \le \Sp e^{\|\A\| \delta^*}\|c\|
\end{equation*}
for all $c\in \R^n$. Hence, $\Phi$, with $|\Theta_i|=1$ for every sampling location $i \in \Omega$, consists of $n$ linearly independent vectors and forms a frame for $\R^n$. By increasing the number of samples per sampling location,  $\Phi$ will remain to be a frame for $\R^n$.

\noindent{\it Proof of Corollary \ref{cor:staterandom}:} First, for each $i \in \Omega$, consider $|\Theta_i|=1$ and denote $\Theta_i=\{t_i\}$. Moreover, we define
\begin{align}
\theta=[t_i]_{i=1,\dots,n}.
\end{align}
Now, we denote $\Sb(\theta)$ to be the frame matrix corresponding to family of vectors $(\ref{eq:nonuniform_complicated})$, where the sampling times have been gathered in $\theta \in \R^n$. Based on Theorem \ref{theorem:statesampling}, there exist a $\theta$ for which the real analytic function $F(\theta):~\R^n  \rightarrow \R$ defined by
$$
F(\theta):=\mathrm{det}(\Sb(\theta)),
$$
is nonzero. Thus, measure of points $\theta \in [0,\tau]^{n}$ for which $F(\theta)$ vanishes is zero (e.g. see \cite{mityagin2015zero}), and independent and uniform sampling of the time stamps in $[0,\tau]^{n}$ gives a frame with probability one.  If we increase $\Theta_i$ beyond $1$ probability of getting a frame is  $1$. 


\noindent{\it Proof of Theorem \ref{theorem:sparse}:}
We build up our proof based on  some of the steps taken in the proof of Theorem 5 in \cite{batson2013spectral}.  Using the leverage scores  \eqref {resistent.def}, one can verify that
\begin{align}\label{eq:spectralbound_first}
{\bf 0} \Sp \prec \Sp (1-\epsilon)  \Sp {\bf S} \Sp \preceq \Sp  \Sb_w
\end{align}
with probability at least $1/2$, where
\[ \Sb_w \Sp = \Sp  \sum_{\phi \in \Phi}\Sp w_s(\phi) \Sp \phi \phi^T \Sp = \Sp \sum_{\phi \in \Phi} \Sp \frac{f_{\phi}}{q\pi(\phi)} \Sp \Sp \phi \phi^T. \]
and $f_{\phi}\geq 0$ is the frequency of the times that $\phi$ is sampled (due to sampling with replacement, some vectors can be sampled multiple times). Weights of those $\phi \notin \Phi_s$ are equal to $0$. For an outcome of Algorithm  \ref{alg:sparsification}, one has
\[ {\bf 0} \Sp \prec \Sp {\bf S}_w \Sp \preceq \Sp  \left( \max_{\phi \in \Phi_s} \Sp w_s(\phi) \right)  \Sp  \sum_{\phi \in \Phi_s} \phi \phi^T \]
This implies that ${\bf S}_s = \sum_{\phi \in \Phi_s} \phi \phi^T \succ {\bf 0}$ with probability at least $1/2$. For the first part of our proof, $\Phi_s$ is a frame for $\R^n$ if and only if $\Sb_s \succ {\bf 0}$.

Denote  the analysis operator of $\Phi_s$ by $\T_s$ and set ${\bf W}_s=\mathrm{diag}\big(w_s(\phi)\big)\big|_{\phi \in \Phi_s}$.
Let
\begin{align}\label{eq:noisy_model2}
y_s=\T_s x_0+\xi_s
\end{align}
be a noisy observation vector collected by $\phi\in \Phi_s$, in which
$\xi_s \in \R^m$ is a zero mean  Gaussian measurement noise with independent components and  covariance
$
\mathbb{E}  \left \{ \xi \xi^T  \right \}=\sigma^2 \I$.  In the next step, let us consider an alternative estimator $\tilde x_0$ that is given by
\begin{eqnarray*}
	\tilde{x}_0 
	& = & \big (\T_s^T  {\bf W}_s  \Sp \T_s\big )^{-1} \Sp \T_s^T {\bf W}_s y_s.
\end{eqnarray*}
It is straightforward to verify that this is an unbiased estimator using only the observations corresponding to  $\Phi_s$. Since covariance of noise is $
\mathbb{E}  \left \{ \xi \xi^T  \right \}=\sigma^2 \I
$, the unweighted least-squares estimator gives the optimal estimator $\hat x_0$.  Therefore,
$$
\mathbb{E}\left \{\tilde x_0 \tilde x_0^T\right \} \succeq  \mathbb{E}\left \{\hat x_0 \hat x_0^T \right \}.
$$
This lets us write
\begin{eqnarray} \notag
\rho_d(\Phi_s)^2&=& \sigma^2 \Sp \mathrm{Tr}\left (\mathbb{E}\{\hat x_0 \hat x_0^T\}\right ) \\
\notag &\leq &\sigma^2 \Sp \mathrm{Tr}\left (\mathbb{E}\{\tilde x_0 \tilde x_0^T\}\right ) \\ &  \notag
= & \sigma^2 \Sp \Tr \left (\Sb_w^{-1} \Sp \T^T {\bf W}_s^2 \Sp  \T \Sp \Sb_w^{-1} \right ) \\& = &\sigma^2 \Sp \Tr \left (\Sb_w^{-1} \Sp \hat \Sb_w \Sp \Sb_w^{-1} \right ) \label{eq:somebound1}
\end{eqnarray}
where  $\Sb_w = \T_s^T {\bf W}_s \T_s$ and $\hat\Sb_w := \T_s^T {\bf W}_s^2  \Sp \T_s$. From \eqref{eq:somebound1} and the definition of random variable  $\chi$ in \eqref{eq:omega_rv}, it follows that
\begin{align}\label{eq:somebound2}
\rho_d(\Phi_s)^2 ~\leq~ \sigma^2  \chi \Sp \mathrm{Tr}\big(\Sb_w^{-1}  \Sb_w \Sb_w^{-1} \big)
\end{align}
in which  the middle matrix is replaced by its upper bound as one can show that for every three positive-definite matrices $\mathbf{X}_1,\mathbf{X}_2,\mathbf{X}_3$ with $\mathbf X_1 \succeq \mathbf X_2$, inequality
$
\mathbf{X}_3 \mathbf X_1 \mathbf{X_3} \succeq \mathbf{X}_3 \mathbf X_2 \mathbf{X_3}
$
holds. According to Markov inequality,  the next inequality holds
\begin{align}\label{eq:someboundomega}
\chi \Sp \leq \Sp \dfrac{1}{1-\frac{3}{4}} \Sp \mathbb{E}\{\chi\} \Sp = \Sp 4\bar \chi
\end{align}
with probability at least $3/4$. From \eqref{eq:somebound2} and \eqref{eq:someboundomega}, we get
\begin{align}\label{eq:somebound3}
\rho_d(\Phi_s) ^2 ~\leq ~4 \Sp \sigma^2 \Sp \bar \chi \Sp \Sp \mathrm{Tr} \big(\Sb_w^{-1}\big).
\end{align}
On the other hand, by applying similar steps to the proof of Theorem 5 in \cite{batson2013spectral}, it follows that with probability at least $1/2$ we have
\begin{align}
(1-\epsilon) \Sb \Sp \preceq \Sp \Sb_w \Sp \preceq \Sp (1+\epsilon) \Sb.
\end{align}
By taking inverse, we get \begin{align}\label{eq:somebound4}
(1+\epsilon)^{-1} \Sb^{-1} \Sp \preceq \Sp \Sb_w^{-1} \Sp \preceq   \Sp (1-\epsilon)^{-1} \Sb^{-1}.
\end{align}
If the event described in \eqref{eq:someboundomega} is denoted by $\mathfrak{A}$ and the event described by \eqref{eq:somebound4} is denoted by $\mathfrak{B}$, then
$$
\mathbb{P}(\mathfrak{A} \cap \mathfrak{B}) = \mathbb{P}(\mathfrak{A})+\mathbb{P}(\mathfrak{B})-\mathbb{P}(\mathfrak{A} \cup \mathfrak{B}) \geq \dfrac{3}{4}+\dfrac{1}{2} -1=\dfrac{1}{4}.
$$
Therefore both \eqref{eq:somebound3} and \eqref{eq:somebound4} hold with probability at least $1/4$.
Combining \eqref{eq:somebound3} and \eqref{eq:somebound4}, one arrives at
\begin{align}\label{eq:somebound5}
\rho_d(\Phi_s)^2  \Sp \leq \Sp \dfrac {4 \sigma^2\bar \chi}{1-\epsilon} \Sp \Sp \mathrm{Tr}(\Sb^{-1})= \dfrac {4 \bar \chi}{1-\epsilon} \Sp \rho_d(\Phi) ^2.
\end{align}
Taking the square root from both sides, we get the  desired inequality \eqref{eq:comega_bound_1}.

For the entropy estimation measure, by following almost identical steps, we can show that with probability at least $1/4$ the following inequality holds
\begin{eqnarray*}
	\rho_e(\Phi_s) &= &\det(\log (\Sb_s^{-1})) \\
	& \leq & \log\left (\det\left (\dfrac{4\bar \chi}{1-\epsilon} \Sb^{-1}\right ) \right ) \\
	&=& n\log \left ( \dfrac{4\bar \chi}{1-\epsilon} \right ) +\rho_e(\Phi).
\end{eqnarray*}

\noindent{\it Proof of Corollary \ref{cor:wbound}:}  We can write
\begin{eqnarray*}
	\sum_{\phi \in \Phi } w_s(\phi)^2 \Sp \phi \phi^T & \leq & \sum_{\phi \in \Phi}  w_s(\phi) \Big( \max_{j=1,\dots,|\Phi|} w_s(\phi)\Big)\,  \phi \phi^T   \\
	& = &  \Big( \max_{j=1,\dots,|\Phi|} w_s(\phi)\Big) \sum_{\phi \in \Phi } w_s(\phi) \Sp \phi \phi^T.
\end{eqnarray*}
Therefore, $\max_{\phi \in \Phi}\Sp w_s(\phi)$ is an upper-bound on $\chi$. The rest of the proof follows from the fact that we can replace $\chi$ with  $\max_{\phi \in \Phi}\Sp w_s(\phi)$ in the proof of Theorem \ref{theorem:sparse}.

\noindent{\it Proof of Theorem \ref{eq:perf_bound}:}
 Let us denote $\Sb_j$ to be the frame matrix corresponding to  family of vectors $\Phi_j$ that is resulted from the random partitioning according to  Proposition \ref{prop:KS}. For the least-squares estimation measure, we have
	\begin{eqnarray*}
		\rho_d(\Phi_j) &=& \sigma \sqrt{\mathrm{Tr} \big (\Sb_j^{-1} \big )}  \leq  \dfrac{\sigma}{\sqrt{1-\dfrac{(1+\sqrt{2r})^2}{2}}}\sqrt{\mathrm{Tr} \left (\Sb^{-1} \right )}  \\
		& = &\Big(1-\dfrac{(1+\sqrt{2r})^2}{2}\Big)^{-1/2}  {\rho_d(\Phi)}.
	\end{eqnarray*}
	This proves the first bound.
	
	For the entropy estimation measure, it follows that
	\begin{eqnarray*}
		\rho_e(\Phi_j) & = & \det(\log (\Sb_j^{-1})) \\
		& \leq & \log\left (\det\left (({{1-\dfrac{(1+\sqrt{2r})^2}{2}}})^{-1} \Sb^{-1}\right ) \right ) \\
		&= & - n\log \left ( 1-\dfrac{(1+\sqrt{2r})^2}{2} \right ) +\rho_e(\Phi),
	\end{eqnarray*}
	which proves the second bound.

\noindent{\it Proof of Proposition \ref{prop:updates}:} 
Taking trace from both sides of \eqref{eq:sinv_update}, we get
$$
\rho_d^2(\Phi_{i+1})=\rho_d^2(\Phi_{i})+\dfrac{\sigma^2}{1-\phi^T\Sb_i^{-1}\phi} \mathrm{Tr}\left ( \Sb_i^{-1} \phi\phi^T \Sb_i^{-1}\right ).
$$
Update rule \eqref{eq:rho_update} follows by utilizing the following equation
$$
\mathrm{Tr}\left ( \Sb_i^{-1} \phi\phi^T \Sb_i^{-1}\right )=\left \| \Sb_i^{-1} \phi \right \|^2.
$$
The update rule \eqref{eq:entropy_update} for the entropy estimation measure follows from
\begin{eqnarray*}
	\rho_e(\Phi_{i+1}) &=& \log\left (\mathrm{det}\left (\Sb_{i+1}^{-1}\right )\right ) \\
	& = &
	\log\left (\mathrm{det}\left(\Sb_{i}^{-1}+\dfrac{\Sb_i^{-1} \phi\phi^T \Sb_i^{-1}}{1-\phi^T\Sb_i^{-1}\phi}\right )\right ) \\
	& =& \log\left (\left(1+\dfrac{(\Sb_i^{-1} \phi)^T (\Sb_i^{-1})^{-1} \Sb_i^{-1}\phi}{1-\phi^T\Sb_i^{-1}\phi}\right)\mathrm{det}(\Sb_{i}^{-1})\right) \\
	& =&  \log\left (\left(1+\dfrac{\phi^T\Sb_i^{-1}\phi}{1-\phi^T\Sb_i^{-1}\phi}\right)\mathrm{det}(\Sb_{i}^{-1})\right) \\
	& = &  \log\left (\left(\dfrac{1}{1-\phi^T\Sb_i^{-1}\phi}\right)\mathrm{det}(\Sb_{i}^{-1})\right) \\
	& = & \log(\mathrm{det}(\Sb_{i}^{-1}))+\log\left (\dfrac{ 1}{1-\phi^T\Sb_i^{-1}\phi}\right) \\
	& = & \rho_e(\Phi_i)-\log(1-\phi^T\Sb_i^{-1}\phi).
\end{eqnarray*}
In the third line,  the matrix determinant lemma is applied  \cite{brookes2005matrix}.

\noindent{\it Proof of Theorem \ref{thm:bounds}:} 
First, we prove a more general inequality. Let us consider the class of all estimation measures that have the following  spectral representation
\begin{align}\label{eq:spectral}
\rho(\Phi)=\sum_{i=1}^n \psi(\lambda_i(\Sb)),
\end{align}
for some convex and monotonically  decreasing function $\psi: \R_+ \rightarrow \R$. Since $\psi$ is convex, we apply Jensen's inequality  \cite{hansen2003jensen} and write
\begin{eqnarray}
\dfrac{1}{n}\rho(\Phi) & = & \dfrac{1}{n}\sum_{i=1}^n \psi(\lambda_i(\Sb)) \nonumber\\
& \geq & \psi\left ( \sum_{i=1}^n \dfrac{\lambda_i(\Sb)}{n} \right )=\psi \left( \dfrac{\mathrm{Tr}(\Sb)}{n} \right ).\label{first}
\end{eqnarray}
On the other hand, we have
\begin{eqnarray}
\mathrm{Tr}\left (\Sb\right ) &=& \mathrm{Tr}\Bigg(~\sum_{i=1}^{|\mathfrak{S}|} \phi_i \phi_i^T ~\Bigg)=\sum_{i=1}^{|\mathfrak{S}|} \Sp \mathrm{Tr}\left (\phi_i \phi_i^T \right ) \notag \\
& = & \sum_{i=1}^{|\mathfrak{S}|} \phi_i^T \phi_i= \sum_{i=1}^{|\mathfrak{S}|} \|\phi_i\|^2. \label{Tr}
\end{eqnarray}
Each vector $\phi_i$ corresponds to some time stamp $t$ and some index $j\in \Omega$, i.e.,
$$
\phi_i=\e^{\A^Tt} e_j.
$$
Thus, we use the bound on the matrix exponential to get
$$
\|\phi_i\| \Sp = \Sp \|\e^{\A^Tt} \Sp e_j\| \Sp \leq \Sp \|\e^{\A^Tt}\| \Sp \|e_j\| \Sp \leq \Sp \nu(\A,\mathfrak{S}).
$$
This inequality together with \eqref{Tr} gives us
$$
\mathrm{Tr}\left (\Sb\right ) \Sp \leq \Sp \sum_{i=1}^{|\mathfrak{S}|} \Sp \|\phi_i\|^2 \Sp \leq \Sp \sum_{i=1}^{|\mathfrak{S}|} \Sp \nu(\A,\mathfrak{S}) ^2=\nu(\A,\mathfrak{S})^2 \Sp |\mathfrak{S}|.
$$
Since $\psi$ is  monotonically decreasing, from \eqref{first}, one may conclude that
\begin{equation}\label{eq:jensenbound}
\dfrac{1}{n}\rho(\Phi) \Sp \geq \Sp \psi \left( \dfrac{\mathrm{Tr}(\Sb)}{n} \right ) \Sp \geq \Sp  \psi \left( \dfrac{\nu(\A,\mathfrak{S})^2 \Sp |\mathfrak{S}|}{n} \right ).
\end{equation}
By applying inequality \eqref{eq:jensenbound} to spectral functions $\psi(\lambda)=\lambda^{-1}$  and $\psi(\lambda)=-\log(\lambda)$, we will get  the desired inequalities \eqref{eq:rho2_bound} and \eqref{eq:bound_ent}, respectively.

\noindent{\it Proof of Theorem \ref{thm:Lyap}:}  
Every sampling strategy $\mathfrak{S}$ that satisfies our assumptions also satisfies	
$$
\mathfrak{S} \subset \Omega \times \delta \mathbb{Z}.
$$
Therefore, the frame matrix corresponding to such  sampling strategy satisfies
\begin{eqnarray}
\Sb  & \preceq &\sum_{i \in \Omega} \Sp \Sp \sum_{t \in \delta \mathbb{Z}_+} \e^{\A^T t} \Sp [\mathbf{C}_{\Omega}]_i^T[\mathbf{C}_{\Omega}]_i \Sp\e^{\A t} \\
& = & \sum_{k=0}^\infty \Big( \e^{\A^T \delta} \Big) ^k \mathbf{C}_{\Omega}^T \mathbf{C}_{\Omega} \Big( \e^{\A \delta} \Big) ^k=\mathbf{Q}.\label{eq:matrixinequality}
\end{eqnarray}
The last equality holds for the following reason. Since $\A$ is Hurwitz, $\e^{\A \delta}$ is Schur and   $(\e^{\A  \delta},\mathbf{C}_\Omega)$ is observable according to \eqref{condi-1}. Inequality  \eqref{eq:matrixinequality} is equivalent to
\begin{align}\label{eq:qinvsbinv}
\mathbf{Q}^{-1}  \preceq \Sb^{-1}.
\end{align}
Functions $\sqrt{\mathrm{Tr}(\bf X)}$ and $\det (\log (\bf X))$ are nondecreasing on the cone of positive semi-definite matrices. By apply these  functions to both sides of \eqref{eq:qinvsbinv}, one obtains  the desired inequalities \eqref{ineq-1} and \eqref{ineq-2}.

\bibliography{mybib}

\end{document}